# Two-dimensional Cold Electron Transport for Steep-slope Transistors

Maomao Liu,[1] Hemendra Nath Jaiswal,[1] Simran Shahi,[1] Sichen Wei,[2] Yu Fu,[2] Chaoran Chang,[2] Anindita Chakravarty,[1] Xiaochi Liu,[3] Cheng Yang,[4] Yanpeng Liu,[5] Young Hee Lee,[6] Fei Yao,[2*] and Huamin Li[1*]

[1] *Department of Electrical Engineering, University at Buffalo, The State University of New York, Buffalo, New York 14260, US*

[2] *Department of Materials Design and Innovation, University at Buffalo, The State University of New York, Buffalo, New York 14260, US*

[3] *School of Physics and Electronics, Central South University, Changsha 410083, China*

[4] *School of Physics and Electronics, Shandong Normal University, Jinan 250014, China*

[5] *Institute of Nanoscience, Nanjing University of Aeronautics and Astronautics, Nanjing 210016, China*

[6] *Center for Integrated Nanostructure Physics, Institute for Basic Science, Suwon 16419, Korea*

[*] Author to whom correspondence should be addressed. Electronic address: feiyao@buffalo.edu and huaminli@buffalo.edu

# Abstract

Room-temperature Fermi-Dirac electron thermal excitation in conventional three-dimensional (3D) or two-dimensional (2D) semiconductors generates hot electrons with a relatively long thermal tail in energy distribution. These hot electrons set a fundamental obstacle known as the "Boltzmann tyranny" that limits the subthreshold swing (SS) and therefore the minimum power consumption of 3D and 2D field-effect transistors (FETs). Here, we investigated a novel graphene (Gr)-enabled cold electron injection where the Gr acts as the Dirac source to provide the cold electrons with a localized electron density distribution and a short thermal tail at room temperature. These cold electrons correspond to an electronic cooling effect with the effective electron temperature of ~145 K in the monolayer $MoS_2$, which enable the transport factor lowering and thus the steep-slope switching (across for 3 decades with a minimum SS of 29 mV/decade at room temperature) for a monolayer $MoS_2$ FET. Especially, a record-high sub-60-mV/decade current density (over 1 µA/µm) can be achieved compared to conventional steep-slope technologies such as tunneling FETs or negative capacitance FETs using 2D or 3D channel materials. Our work demonstrates the great potential of 2D Dirac-source cold electron transistor as an innovative steep-slope transistor concept, and provides new opportunities for 2D materials toward future energy-efficient nanoelectronics.

## Main

Fermi-Dirac electron thermal excitation is an intrinsic physical phenomenon that grants a tunable electrical conductivity of semiconductors at room temperature, but on the other hand, it also causes excessive power dissipation in various electron systems [1-3]. Especially in field-effect transistors (FETs), the "Boltzmann tyranny" induced by the thermal excitation of hot electrons sets a fundamental limit in the steepness of the transition slope between off and on states, known as subthreshold swing (SS). To change the current by one order of magnitude at room temperature, the minimum gate voltage is required to be $\partial V_G/\partial(\log_{10}I_D) = (\partial V_G/\partial \Psi_s)[\partial \Psi_s/\partial(\log_{10}I_D)] \sim (k_BT/q)\ln10 \sim 60$ mV, where $V_G$ is the gate voltage, $I_D$ is the drain current, $\Psi_s$ is the channel surface potential, $k_B$, $T$, and $q$ are the Boltzmann constant, temperature, and electronic charge, respectively. Therefore, various types of steep-slope FETs have been proposed to increase the turn-on steepness and overcome the bottleneck to continue minimizing the power consumption, including tunneling FETs (TFETs) [1-3] and negative capacitance FETs (NCFETs) [4-6]. These solutions in principle can achieve a sub-60-mV/decade SS, for example, by lowering the transport factor ($\partial \Psi_s/\partial(\log_{10}I_D)$), via a band-to-band Zener tunneling effect in TFETs [7-10]; or by reducing the body factor ($\partial V_G/\partial \Psi_s$) via a ferroelectric gate layer with the negative differential capacitance in NCFETs [11-13]. However, there are also critical challenges for these technologies to overcome, such as the low current density and source/drain asymmetry in TFETs [14-16] as well as the lack of fundamental understanding in NCFETs [17-20].

With the rise of graphene (Gr), two dimensional (2D) van der Waals (vdW) layered materials have been explored as promising material candidates for future energy-efficient nanoelectronics due to the natural quantum confinement in an atomically thin body [21, 22]. $MoS_2$ as one of the most representative semiconducting transition metal dichalcogenides (TMDs) acts as an excellent channel material for nano-scale transistors by offering ideal electrostatic tunability, an appropriate direct bandgap, and moderate carrier mobility, etc [23-25]. The $MoS_2$-based TFETs [7-10] and NCFETs [11-13] have been demonstrated with the sub-60-mV/decade SS. However, neither of them can provide the sub-60-mV/decade current density to be higher than 1 µA/µm which is one key metric for a logic transistor to benefit from the steep slope [15, 16, 26].

In this work, we demonstrated a novel Gr-enabled Dirac-source "cold" electron injection which possesses a more localized electron density distribution and a shorter thermal tail, compared to the conventional normal-source hot electron injection in 3D or 2D semiconductors. The cold electron injection has been implemented in a monolayer $MoS_2$ FET to introduce an electronic cooling effect which can lower the transport factor and thus enable an outstanding steep-slope switching (across for 3 decades with a minimum SS of 29 mV/decade at room temperature), an excellent on/off ratio (~$10^7$), novel and strong on-current saturation (~10 µA/µm), and especially a record-high sub-60-mV/decade current density (over 1 µA/µm) compared to any TFETs or NCFETs using 2D or 3D channel materials. The effective electron temperature was extracted to be ~145 K, based on the energy distribution of the cold electrons at room temperature. Our work presents the 2D Dirac-source cold electron FET as an innovative steep-slope transistor concept to achieve the sub-60-mV/decade switching, and provides new opportunities for 2D materials toward future energy-efficient nanoelectronics.

**Graphene-enabled 2D "cold" electron transport.** To illustrate the underlying mechanism and the fundamental advantages of the Gr-enabled Dirac-source cold electron transport, we compare it with the conventional normal-source hot electron transport in 3D and 2D semiconductors such as Si and monolayer $MoS_2$. The electron density (*n*) depends on the energy (*E*), and is defined as the product of the Fermi-Dirac distribution function $f(E) = 1/\{1+\exp[(E-$ $E_F)/k_BT]\}$ and the density of states (DOS), where $E_F$ is the Fermi energy. Here we use the subscript of 3D, 2D, and Gr to distinguish the properties for 3D semiconductors, 2D semiconductors, and monolayer Gr, respectively. Because DOS is a parabolic function of $E$ in 3D semiconductors as $DOS_{3D}(E) \sim (E–E_C)^{1/2}$, the corresponding $n(E)$ can be expressed as $n_{3D}(E) \sim (E–E_C)^{1/2}\exp[(E_F–E)/k_BT]$ which follows a sub-exponential decay, as shown in **Fig. 1(a)**, where $E_C$ is the minimum conduction band edge. Similarly, since DOS in 2D semiconductors is constant ($DOS_{2D}(E) \sim (E–E_C)^0$), the corresponding $n(E)$ possesses an exponential decay described as $n_{2D}(E) \sim \exp[(E_F–E)/k_BT]$, as shown in **Fig. 1(b)**. Assuming $E_F = 0$ eV, the exponential decay of $n_{2D}(E)$ also follows the Maxwell-Boltzmann distribution function known as $n(E) \sim \exp(–E/k_BT)$. Both $n_{3D}(E)$ and $n_{2D}(E)$ present a relatively long Boltzmann thermal tail at room temperature, and these tails extend to infinity at the non-zero temperature in principle, due to the Fermi-Dirac electron thermal excitation. Considering a conventional FET structure where a gate-controlled potential barrier ($\phi_b$) is established within the channel, the drain current is attributed to the thermionic injection of the electrons with $E$ higher than $\phi_b$. Due to the long Boltzmann thermal tail as $E$ increases, the minimum SS for 3D and 2D semiconductor channels suffers from the "hot" electron injection and has a limit of 60 mV/decade at room temperature.

As a comparison, the monolayer Gr has a linear energy dispersion near the Dirac point, and its DOS is a linear function of $E$ as $DOS_{Gr}(E) \sim E_{Dirac}–E$ where $E_{Dirac}$ is the energy at the Dirac point. The corresponding $n(E)$ of the monolayer Gr can be expressed as $n_{Gr}(E) \sim (E_{Dirac}–E)/\{1+\exp[(E–E_F)/k_BT]\} \sim (E_{Dirac}–E)\exp[(E_F–E)/k_BT] \sim (E_{Dirac}–E)n_{2D}(E)$. Thus, when $E_{Dirac}–E < 1$, $n_{Gr}(E)$ always shows a super-exponential decay compared to $n_{2D}(E)$. The Boltzmann thermal tail in Gr is relatively short, because it is no longer infinite but terminated right at $E_{Dirac}$ even at room temperature, as shown in **Fig. 1(c)**. By exploiting the monolayer Gr as the cold electron source in a heterojunction structure, for example, a Gr/MoS$_2$ heterobilayer, $\phi_b$ can be created at the interface and tuned by the gate electrostatically. Compared to the hot electron injection in 3D and 2D semiconductors, the cold electron injection from Gr possesses a more localized distribution near $E_F$ and a shorter thermal tail which can be cut off more effectively by $\phi_b$, giving rise to a faster switching to break the SS limit in principle [27, 28].

To provide a quantitative comparison, we assume $n(E_F)$ is the electron density at $E = E_F$, and calculate the normalized carrier density, i.e., $n(E)/n(E_F)$ for the monolayer Gr and 2D semiconductors, as shown in **Fig. 2(a)**. Here Gr is p-type doped and $E_{Dirac}$–$E_F$ ranges from 0.1 to 0.5 eV. Assuming that a gate-controlled $\phi_b$ increases from 0 to 0.1 eV, $n(E)/n(E_F)$ in 2D semiconductors is reduced from 1 to 0.021. A modulation efficiency, defined as $\eta = 1 - n(E)/n(E_F)$, suggests that 97.9% electrons above $E_F$ is cut off by $\phi_b$, as shown in **Fig. 2(b)**. In contrast, for the p-type doped Gr where $E_{Dirac} = 0.1$ eV, all the electrons above $E_F$ are completely halted by $\phi_b$, giving rise to $\eta = 1$. As the doping level increases in Gr, $\eta$ slightly decreases but is still higher than that of 2D semiconductors even at $E_{Dirac} = 0.5$ eV. It is worth mentioning that 2D semiconductors in this comparison are considered as an extreme case in which $E_F$ aligns with $E_C$ as $E_F = E_C = 0$ eV. For a more practical situation, $E_C$ is usually higher than $E_F$ in 2D semiconductors, and the corresponding $\eta$ would be significantly lowered compared to that in Gr. For example, assuming $E_F = 0$ eV, $E_C = 0.1$ eV, and $\phi_b$ increases from 0.1 to 0.2 eV, $\eta$ is anticipated to be only 2.1% for 2D semiconductors.

Furthermore, when we compare the energy distribution of $n(E)/n(E_F)$ in Gr at room temperature with that in 2D semiconductors at low temperature, it is intriguing to see that they are comparable in certain regimes, as shown in **Fig. 2(c)**. To extract the effective temperature of these cold electrons in 2D semiconductors, we start with the ratio of electron density between Gr and 2D semiconductors, i.e., $n_{Gr}(E)/n_{2D}(E) \sim E_{Dirac}-E$, which strongly depends on the doping level in Gr (see **Fig. 2(a)**). Considering $n_{Gr}(E) \sim (E_{Dirac}-E)\exp[(E_F-E)/k_B T_{Gr}]$ and $n_{2D}(E) \sim \exp[(E_F-E)/k_B T_{2D}]$ where $T_{Gr}$ and $T_{2D}$ are the electron temperature in Gr and 2D semiconductors, respectively, the relationship between $T_{Gr}$ and $T_{2D}$ can be extracted as $T_{2D} = [1/T_{Gr}+k_B\ln(E_{Dirac}-E)/(E_F-E)]^{-1}$ when $n_{Gr}(E)$ equals to $n_{2D}(E)$. At $E_F = 0$ eV, $T_{2D}$ can be further simplified as $T_{2D} = [1/T_{Gr}-k_B\ln(E_{Dirac}-E)/E]^{-1}$. Therefore, when $E_{Dirac} > E > E_F$, the electrons in Gr at room temperature ($T_{Gr} = 300$ K) can serve as the cold electrons in 2D semiconductors at low temperature, as shown in **Fig. 2(d)**. The cooling efficiency, defined as $\gamma = 1-T_{2D}/T_{Gr}$, can be calculated as well, as shown in **Fig. 2(e)**. As $T_{Gr}$ rises, $\gamma$ increases accordingly, suggesting an enhanced capability of suppressing the electron thermal excitation in 2D semiconductors. One can imagine that, by implementing this Gr-enabled cold electron injection in 2D FET devices such as the monolayer $MoS_2$ FET, an electronic cooling effect can be introduced in the electron transport. These cold electrons can equivalently lower the transport factor (($k_B T/q$)ln10) and thus create a new approach to achieve the energy-efficient steep-slope transistors, in addition to the conventional technique using the band-to-band Zener tunneling.

**Sub-60-mV/decade switching of 2D $MoS_2$ Dirac-source cold electron FET.** Here we designed a 2D steep-slope Dirac-source FET (DSFET) using the monolayer Gr and $MoS_2$ as the cold-electron source and semiconductor channel, respectively, and experimentally demonstrated the electronic cooling effect and thus the outstanding sub-60-mV/decade switching performance at room temperature. Schematic of a 2D $MoS_2$ DSFET and the microscopic images of a device are illustrated in **Fig. 3(a-c)**. The details of material characterization and device fabrication are provided in **Supporting Information Note 1 and 2**. In brief, a Gr/$MoS_2$ heterobilayer structure was formed, followed by electrode patterning and deposition. Their electronic properties were investigated through various back-gate transistor configurations including Gr FET, MoS$_2$ FET, and Gr/MoS$_2$ FET. The whole device was passivated by a thin Al$_2$O$_3$ layer, except for the gating windows which were opened in the overlapping area and the top-gate electrode. Room-temperature ionic liquid (diethylmethyl(2-methoxyethyl) ammonium bis(trifluoromethylsulfonyl)imide, or DEME-TFSI) was drop-casted on the device to connect the top-gate electrode to the exposed Gr/MoS$_2$ heterobilayer area, and provide a localized high-efficiency gating through an electric double layer (EDL) effect.

For device characterization, drain current density ($J_D$) as a function of drain voltage ($V_D$), back-gate voltage ($V_{BG}$), and top-gate voltage ($V_{TG}$) were measured at room temperature. First, we use the Si back gate to modulate the carrier transport. Ohmic contacts are confirmed through the $J_D$-$V_D$ output characteristics in the back-gate Gr FET, MoS$_2$ FET, and Gr/MoS$_2$ FET, as shown in **Supporting Information Note 3**. A comparison of the $J_D$-$V_{BG}$ transfer characteristics is summarized in **Fig. 3(d)** and **Supporting Information Note 4**. The Gr FET shows the Dirac point at about –60 V with a weak gate modulation (an on/off ratio less than 2) due to its zero bandgap. The MoS$_2$ FET possesses a typical electron transport branch with an on/off ratio near $10^6$. Similarly, the Gr/MoS$_2$ FET is dominated by electron transport and an intermediate on/off ratio about $10^4$ is obtained. Because the hot carrier injection dominates in all three devices under the back gating, their SS is still constrained by the thermionic limit. Then, we exploit the top gate through the localized EDL effect to precisely control the carrier transport only at the Gr/MoS$_2$ interface, and measure the $J_D$-$V_{TG}$ transfer characteristics for the top-gate MoS$_2$ DSFET at different $V_{BG}$ levels, as shown in **Fig. 3(e)** and **Supporting Information Note 5-8**. It is intriguing to see that a sub-60-mV/decade switching at room temperature can be obtained at $V_{BG}$ = –80 V, and the minimum SS value can be obtained as 49 mV/decade in a forward sweep and 29 mV/decade in a backward sweep, as shown in **Fig. 3(f)**. By extracting the SS as a function of $J_D$, the sub-60-mV/decade switching sustains for about one decade in the forward sweep and about three decades in the backward sweep, as shown in **Fig. 3(g)** and **(h)**. As $V_{BG}$ continues increasing, the minimum SS rises and stabilizes at about 210 mV/decade in the forward sweep and about 90 mV/decade in the backward sweep, as shown in **Fig. 3(i)**. Besides, novel and strong current saturation occurs in both the forward and backward sweeps when the MoS$_2$ DSFET is operated at on state (see **Fig. 3(e)**), where the on-current density is nearly constant (~ 10 μA/μm at $V_D = 0.1$ V) across a wide sweeping range (~ 5 V in the forward sweep and ~ 3 V in the backward sweep). Such unique independence of the gate voltage is not obtained in either the back-gate Gr FET, MoS$_2$ FET, or Gr/MoS$_2$ FET (see **Fig. 3(d)**).

To better understand the steep-slope switching mechanism of the 2D MoS$_2$ DSFET and its dependence on the top-gate and back-gate electric fields, an energy band diagram along the channel is illustrated in **Fig. 4(a)**. The electron transport path from source to drain can be divided into three regions: (i) $V_{BG}$-controlled Gr Dirac source region along the in-plane direction (*x*-axis), (ii) $V_{TG}$-controlled Gr/MoS$_2$ heterobilayer region along the out-of-plane direction (*y*-axis), and (iii) $V_{BG}$-controlled MoS$_2$ channel region along the in-plane direction (*x*-axis). To enable the cold electron injection into the n-type MoS$_2$ channel, the Gr source region is doped into p-type by applying a constant $V_{BG}$ ($V_{BG} < V_{Dirac}$ where $V_{Dirac}$ is the gate voltage at the Dirac point). When the MoS$_2$ DSFET is at the off state, the applied $V_{TG}$ turns the Gr into p-type in the heterobilayer region. $E_F$ in the p-type Gr and $E_C$ in the n-type MoS$_2$ create a large $\phi_b$ at the Gr/MoS$_2$ interface to prevent the electron injection. As $V_{TG}$ increases, both $E_{Dirac}$ in Gr and $E_C$ in MoS$_2$ in the heterobilayer region are lowered, resulting in a transition of Gr from p-type to n-type as well as a reduced $\phi_b$. Thus the MoS$_2$ DSFET is operated in the subthreshold regime, and the current increases due to the thermionic injection of the hot electrons which have higher $E$ over $\phi_b$. As $V_{TG}$ continues increasing, $E_C$ in MoS$_2$ becomes lower than $E_F$ in Gr. Thus, an energy window defined as $E_{Dirac}$–$\phi_b$ is opened for the cold electron injection from the Gr source to the MoS$_2$ channel. With the expansion of the injection window, $J_D$ increases accordingly and eventually reaches the maximum at the on state. Although the carrier transport during the subthreshold regime is still dominated by the thermionic emission over $\phi_b$, the sub-60-mV/decade switching becomes possible due to the localized electron density distribution and the short Boltzmann thermal tail of the cold electrons from Gr. At the on state, because of a degenerate doping from the EDL effect [29, 30], the injection window is maximized and stabilized with the minimum $\phi_b$, giving rise to the strong and steady on-current saturation even when $V_{TG}$ continues increasing.

In addition to the sub-60-mV/decade steep slope and the strong on-current saturation, another intriguing feature of the 2D MoS$_2$ DSFET is the "double minima" in the SS during the operation (see **Fig. 3(g)** and **(h)**, and **Supporting Information Note 8**). In the normal FET operation, the SS-$J_D$ characteristics usually show a single valley, and the minimum SS is attributed to the most efficient thermionic emission of the hot electrons over $\phi_b$ in the subthreshold regime. Whereas in the 2D MoS$_2$ DSFET operation, in addition to the minimum SS induced by the normal-source hot electron injection, the second valley with a sub-60-mV/decade SS occurs, owing to the Dirac-source cold electron injection. Because both the hot and cold electron injections are dominated by the thermionic emission (see **Supporting Information Note 9**), their currents can be described by the Landauer-Büttiker formula at the ballistic transport limit [27, 28, 31], and the SS can be plotted as a function of $\phi_b$–$E_{Dirac}$ for a variety of $C$, as shown in **Fig. 4(b)** and **Supporting Information Note 10**. Here $C = \partial\phi_b/\partial(qV_{TG})$ is the reciprocal body factor describing the variation of $\phi_b$ as a function of $V_{TG}$ for 2D heterojunction, and it ranges from 0 to 1. When $\phi_b > E_{Dirac}$, the Gr serves as the normal source to provide the hot electrons, and the SS is always larger than 60 mV/decade at room temperature. As $V_{TG}$ increases, $\phi_b$ is close to $E_{Dirac}$, and the SS approaches to infinity in principle. When $\phi_b < E_{Dirac}$, the Gr acts as the Dirac source that enables the cold electron injection and allows a faster switching by breaking the SS limit. It is noted that, as $C$ decreases, the steep-slope energy window for the cold electron injection is also reduced, but the sub-60-mV/decade switching is still feasible even at $C = 0.2$. Assuming the EDL capacitance in the ionic liquid is 1 μF/cm$^2$ [32], the Fermi level shift ($E_F$–$E_{Dirac}$) of Gr under the top gating can be estimated [33], and the SS as a function of $V_{TG}$–$V_{Dirac}$ is obtained, as shown in **Fig. 4(c)**. As $V_{TG}$ increases, the hot electron injection and cold electron injection are predicted in succession, which are qualitatively consistent with the experimental data obtained from both forward and backward sweeps.

Owing to the ultrahigh capacitance of EDL, the body factor is assumed to be unity in the 2D MoS$_2$ DSFET. This is also evidenced by the near-60-mV/decade achieved in the hot carrier injection regime. Therefore, the SS is predominated by the transport factor, as shown in **Fig. 5(a)**. Because of the novel electronic cooling effect from the cold electron injection, the transport factor can be lowered than 60 mV/decade at room temperature, and the effective temperature ($T_{eff}$) of the cold electrons can be estimated from the SS. Based on the minimum SS of 29 mV/decade at room temperature, the lowest $T_{eff}$ can be calculated as $T_{eff} = SSq/(k_B\ln10) = $ ~145 K, which is more than 50% reduced.

Finally, we benchmark our 2D MoS$_2$ DSFET with other state-of-the-art beyond-CMOS technologies. Compared to the 14 nm Si FinFET CMOS technology [34], the 2D MoS$_2$ DSFET shows the near-60-mV/decade SS at the beginning of the subthreshold regime owing to the hot electron injection. Then it switches sequentially to the cold electron injection with the sub-60-mV/decade SS until the end of the subthreshold regime, as shown in **Fig. 5(b)**. Although the strongly saturated on-current density is lower than the Si technology, it can be further improved in principle by increasing the doping level of the Gr source and thus enlarging the energy window for the cold electron injection. The sub-60-mV/decade SS of 2D $MoS_2$ DSFET is also plotted as a function of $J_D$ in comparison with other steep-slope technologies, including the TFETs [7, 35-42], NCFETs [12, 43, 44], and one-dimensional (1D) DSFETs [27] based on a variety of channel materials, as shown in **Fig. 5(c)**. The 2D $MoS_2$ DSFET possesses the ultimately thin channel (~0.65 nm for monolayer $MoS_2$) compared to the conventional bulk channel materials such as Si, Ge, and InGaAs, etc. Furthermore, the 2D $MoS_2$ DSFET shows a superior steep-slope current density (~4 µA/µm) which is the highest compared to any TFET and NCFET technologies based on 2D or 3D channel materials so far. Such a high sub-60-mV/decade current density in the 2D $MoS_2$ DSFET is attributed to the unique operation mechanism. The cold electron injection dominates until the end of the subthreshold regime where the relevant subthreshold current is high and close to the saturated on-current density. Whereas the other steep-slope technologies such as the TFETs dominate only at the beginning of the subthreshold regime where the subthreshold current is relatively low. It is also worth mentioning that the sub-60-mV/decade current density of 2D $MoS_2$ DSFET only remains inferior to that of 1D carbon nanotube (CNT) DSFETs, which can be attributed to the employment of high-density CNT arrays as the channel in their case. Here we chose the monolayer $MoS_2$ as the ultimately thin channel simply because it is the most representative 2D semiconducting material. One can easily expect that the richness of 2D material family will further enable the performance-boosting of the 2D DSFETs, for example, by exploiting other high-mobility 2D semiconductors such as black phosphorus.

  In conclusion, we investigated the novel Gr-enabled cold electron injection and thus the electronic cooling effect in monolayer MoS$_2$, and demonstrated the first steep-slope 2D MoS$_2$ DSFET which showed the outstanding sub-60-mV/decade SS (across for 3 decades with the minimum SS of 29 mV/decade at room temperature), the excellent on/off ratio (~10$^7$), the novel and strong on-current saturation (~10 μA/μm across 5 V sweeping range at $V_D$ = 0.1 V), and more importantly, the record-high steep-slope current density (over 1 μA/μm) than any 2D- or 3D-channel-based TFETs or NCFETs. The double-minima SS in the subthreshold regime of the 2D MoS$_2$ DSFET was interpreted, which was attributed to the conventional hot electron injection with the near-60-mV/decade SS and the cold electron injection ($T_{\text{eff}}$ of ~145 K at room temperature) with the sub-60-mV/decade SS. Our work demonstrated the 2D DSFET as a new type of emerging steep-slope transistor concept for beyond-CMOS technology, and revealed innovative opportunities for future nanoelectronics based on 2D layered materials and their vdW heterostructures.

## Data availability

The data that support the findings of this study are available from the corresponding author on reasonable request.

## References

[1]  A. M. Ionescu and H. Riel, "Tunnel field-effect transistors as energy-efficient electronics switches," *Nature*, vol. 479, pp. 329, 2011.

[2]  A. C. Seabaugh and Q. Zhang, "Low-voltage tunnel transistors for beyond CMOS logic," *Proc. IEEE*, vol. 98, pp. 2095, 2010.

[3]  D. Jena, "Tunneling transistors based on graphene and 2-D crystals," *Proc. IEEE*, vol. 101,

pp. 1585, 2013.

[4] A. I. Khan, K. Chatterjee, B. Wang, S. Drapcho, L. You, C. Serrao, S. R. Bakaul, R. Ramesh, and S. Salahuddin, "Negative capacitance in a ferroelectric capacitor," *Nat. Mater.*, vol. 14, pp. 182, 2015.

[5] J. C. Wong and S. Salahuddin, "Negative capacitance transistors," *Proc. IEEE*, vol. 107, pp. 49, 2019.

[6] X. Wang, P. Yu, Z. Lei, C. Zhu, X. Cao, F. Liu, L. You, Q. Zeng, Y. Deng, C. Zhu, J. Zhou, Q. Fu, J. Wang, Y. Huang, and Z. Liu, "Van der Waals negative capacitance transistors," *Nat. Commun.*, vol. 10, no. 3037, 2019.

[7] D. Sarkar, X. Xie, W. Liu, W. Cao, J. Kang, Y. Gong, S. Kraemer, P. M. Ajayan, and K. Banerjee, "A subthermionic tunnel field-effect transistor with an atomically thin channel," *Nature*, vol. 526, pp. 91, 2015.

[8] Y. Balaji, Q. Smets, C. J. L. De La Rosa, A. K. A. Lu, D. Chiappe, T. Agarwal, D. H. C. Lin, C. Huyghebaert, I. Radu, D. Mocua, and G. Groeseneken, "Tunneling transistors based on $MoS_2$/$MoTe_2$ van der Waals heterostructures," *IEEE J. Electron Device Soc.*, vol. 6, pp. 1048, 2018.

[9] X. Liu, D. Qu, H. Li, I. Moon, F. Ahmed, C. Kim, M. Lee, Y. Choi, J. H. Cho, J. C. Hone, and W. J. Yoo, "Modulation of quantum tunneling via a vertical two-dimensional black phosphorus and molybdenum disulfide p-n junction," *ACS Nano*, vol. 11, pp. 9143, 2017.

[10] T. Roy, M. Tosun, X. Cao, H. Fang, D.-H. Lien, P. Zhao, Y.-Z. Chen, Y.-L. Chueh, J. Guo, and A. Javey, "Dual-gated $MoS_2$/$WSe_2$ van der Waals tunnel diodes and transistors," *ACS Nano*, vol. 9, pp. 2071, 2015.

[11] M. Si, C.-J. Su, C. Jiang, N. J. Conrad, H. Zhou, K. D. Maize, G. Qiu, C.-T. Wu, A. Shakouri,

M. A. Alam, and P. D. Ye, "Steep-slope hysteresis-free negative capacitance MoS$_2$ transistors," *Nat. Nanotech.*, vol. 13, pp. 24, 2018.

[12] Z. Yu, H. Wang, W. Li, S. Xu, X. Song, S. Wang, P. Wang, P. Zhou, Y. Shi, Y. Chai, and X. Wang, "Negative capacitance 2D MoS$_2$ transistors with sub-60mV/dec subthreshold swing over 6 orders, 250 µA/µm current density, and nearly-hysteresis-free," *Proc. IEDM Tech. Dig.*, pp. 577, 2017.

[13] F. A. McGuire, Y.-C. Lin, K. Price, G. B. Rayner, S. Khandelwal, S. Salahuddin, and A. D. Franklin, "Sustained sub-60 mV/decade switching via the negative capacitance effect in MoS$_2$ transistors," *Nano Lett.*, vol. 17, pp. 4801, 2017.

[14] U. E. Avci, D. H. Morris, and I. A. Young, "Tunnel field-effect transistors: Prospects and challenges," *IEEE J. Electron Dev. Soc.*, vol. 3, pp. 88, 2015.

[15] A. Seabaugh, C. Alessandri, M. A. Heidarlou, H.-M. Li, L. Liu, H. Lu, S. Fathipour, P. Paletti, P. Pandey, and T. Ytterdal, "Steep slope transistors: Tunnel FET and beyond," *The 46[th] European Solid-State Device Research Conference (ESSDERC)*, pp. 349, 2016.

[16] W. G. Vandenberghe, A. S. Verhulst, B. Sorée, W. Magnus, G. Groeseneken, Q. Smets, M. Heyns, and M. V. Fischetti, "Figure of merit for and identification of sub-60 mV/decade devices," *Appl. Phys. Lett.*, vol. 102, pp. 013510, 2013.

[17] M. Hoffmann, S. Slesazeck, U. Schroeder, and T. Mikolajick, "What's next for negative capacitance electronics?" *Nat. Electron.*, vol. 3, pp. 504, 2020.

[18] W. Cao and K. Banerjee, "Is negative capacitance FET a steep-slope logic switch?" *Nat. Commun.*, vol. 11, no. 196, 2020.

[19] M. A. Alam, M. Si, and P. D. Ye, "A critical review of recent progress on negative capacitance field-effect transistors," *Appl. Phys. Lett.*, vol. 14, no. 090401, 2019.

[20] S. Chang, U. E. Avci, D. E. Nikonov, and I. A. Young, "A thermodynamic perspective of negative-capacitance field-effect transistors," *IEEE J. Explor. Solid-State Computat.*, vol. 3, pp. 56, 2017.

[21] K. S. Novoselov, A. Mishchenko, A. Carvalho, and A. H. Castro Neto, "2D materials and van der Waals heterostructures," *Science*, vol. 353, pp. 461, 2016.

[22] Q. H. Wang, K. Kalantar-Zadeh, A. Kis, J. N. Coleman, and M. S. Strano, "Electronics and optoelectronics of two-dimensional transition metal dichalcogenides," *Nat. Nanotech.*, vol. 7, pp. 699, 2012.

[23] Y. Yoon, K. Ganapathi, and S. Salahuddin, "How good can monolayer $MoS_2$ transistors be?" *Nano Lett.*, vol. 11, no. 9, pp. 3768, 2011.

[24] X. Li, L. Yang, M. Si, S. Li, M. Huang, P. Ye, and Y. Wu, "Performance potential and limit of $MoS_2$ transistors," *Adv. Mater.*, vol. 27, pp. 1547, 2015.

[25] K. Alam and R. K. Lake, "Monolayer $MoS_2$ transistors beyond the technology road map," *IEEE Trans. Electron Dev.*, vol. 59, no. 12, pp. 3250, 2012.

[26] G. V. Resta, A. Leonhardt, Y. Balaji, S. De Gendt, P.-E. Gaillardon, and G. De Micheli, "Devices and circuits using novel 2-D materials: A perspective for future VLSI systems," *IEEE Trans. Very Large Scale Integr. (VLSI) Syst.*, vol. 27, pp. 1486, 2019.

[27] C. Qiu, F. Liu, L. Xu, B. Deng, M. Xiao, J. Si, L. Lin, Z. Zhang, J. Wang, H. Guo, H. Peng, and L.-M. Peng, "Dirac-source field-effect transistors as energy-efficient, high-performance electronic switches," *Science*, vol. 361, pp. 387, 2018.

[28] F. Liu, C. Qiu, Z. Zhang, L.-M. Peng, J. Wang, and H. Guo, "Dirac electrons at the source: breaking the 60-mV/decade switching limit," *IEEE Trans. Electron Dev.*, vol. 65, pp. 2736, 2018.

[29] P. Paletti, R. Yue, C. Hinkle, S. K. Fullerton-Shirey, and A. Seabaugh, "Two-dimensional electric-double-layer Esaki diode," *npj 2D Mater. Appl.,* vol. 3, no. 19, 2019.

[30] C. Alessandri, S. Fathipour, H. Li, I. Kwak, A. Kummel, M. Remskar, and A. Seabaugh, "Reconfigurable electric double layer doping in an MoS$_2$ nanoribbon transistor," *IEEE Trans. Electron Dev.*, vol. 64, no. 12, pp. 5217, 2017.

[31] L. Britnell, R. V. Gorbachev, R. Jalil, B. D. Belle, F. Schedin, A. Mishchenko, T. Georgiou, M. I. Katsnelson, L. Eaves, S. V. Morozov, N. M. R. Peres, J. Leist, A. K. Geim, K. S. Novoselov, and L. A. Ponomarenko, "Field-effect tunneling transistor based on vertical graphene heterostructures," *Science*, vol. 335, pp. 947, 2012.

[32] Y. Zhang, J. Ye, Y. Matsuhashi, and Y. Iwasa, "Ambipolar MoS$_2$ thin flake transistors," *Nano Lett.*, vol. 12, pp. 1136, 2012.

[33] F. Wang, Y. Zhang, C. Tian, C. Girit, A. Zettl, M. Crommie, and Y. R. Shen, "Gate-variable optical transitions in graphene," *Science*, vol. 320, pp. 206, 2008.

[34] S. Natarajan, M. Agostinelli, S. Akbar, M. Bost, A. Bowonder, V. Chikarmane, S. Chouksey, A. Dasgupta, K. Fischer, Q. Fu, T. Ghani, M. Giles, S. Govindaraju, R. Grover, W. Han, D. Hanken, E. Haralson, M. Haran, M. Heckscher, R. Heussner, P. Jain, R. James, R. Jhaveri, I. Jin, H. Kam, E. Karl, C. Kenyon, M. Liu, Y. Luo, R. Mehandru, S. Morarka, L. Neiberg, P. Packan, A. Paliwal, C. Parker, P. Patel, R. Patel, C. Pelto, L. Pipes, P. Plekhanov, M. Prince, S. Rajamani, J. Sandford, B. Sell, S. Sivakumar, P. Smith, B. Song, K. Tone, T. Troeger, J. Wiedemer, M. Yang, and K. Zhang, "A 14nm logic technology featuring 2$^{nd}$-generation FinFET, air-gapped interconnects, self-aligned double patterning and a 0.0588μm$^2$ SRAM cell size," *Proc. IEDM Tech. Dig.*, pp. 371, 2014.

[35] G. Dewey, B. Chu-Kung, J. Boardman, J. M. Fastenau, J. Kavalieros, R. Kotlyar, W. K. Liu,

D. Lubyshev, M. Metz, N. Mukherjee, P. Oakey, R. Pillarisetty, M. Radosavljevic, H. W. Then, and R. Chau, "Fabrication, characterization, and physics of III-V heterojunction tunneling field effect transistors (H-TFET) for steep sub-threshold swing," *Proc. IEDM Tech. Dig.*, pp. 785, 2011.

[36] R. Gandhi, Z. Chen, N. Singh, K. Banerjee, and S. Lee, "CMOS-compatible vertical-silicon-nanowire gate-all-around p-type tunneling FETs with ≤50-mV/decade subthreshold swing," *IEEE Electron Dev. Lett.*, vol. 32, pp. 1504, 2011.

[37] K. Jeon, W.-Y. Loh, P. Patel, C. Y. Kang, J. Oh, A. Bowonder, C. Park, C. S. Park, C. Smith, P. Majhi, H.-H. Tseng, R. Jammy, T.-J. K. Liu, and C. Hu, "Si tunnel transistors with a novel silicided source and 46mV/dec swing," *Proc. VLSI Symp. Tech. Dig., pp.* 121, 2010.

[38] S. H. Kim, H. Kam, C. Hu, and T.-J. K. Liu, "Germanium-source tunnel field effect transistors with record high $I_{ON}/I_{OFF}$," *Proc. VLSI Symp. Tech. Dig.*, pp. 178, 2009.

[39] L. Knoll, Q.-T. Zhao, A. Nichau, S. Trellenkamp, S. Richter, A. Schafer, D. Esseni, L. Selmi, K. K. Bourdelle, and S. Mantl, "Inverters with strained Si nanowire complementary tunnel field effect transistors," *IEEE Electron Dev. Lett.*, vol. 34, pp. 813, 2013.

[40] Y. Lu, S. Bangsaruntip, X. Wang, L. Zhang, Y. Nishi, and H. Dai, "DNA functionalization of carbon nanotubes for ultrathin atomic layer deposition of high κ dielectrics for nanotube transistors with 60 mV/decade switching," *J. Am. Chem. Soc.*, vol. 128, pp. 3518, 2006.

[41] T. Mori, T. Yasuda, K. Fukuda, Y. Morita, S. Migita, A. Tanabe, T. Maeda, W. Mizubayashi, S.-I. O'uchi, Y. Liu, M. Masahara, N. Miyata, and H. Ota, "Unexpected equivalent-oxide-thickness dependence of the subthreshold swing in tunnel field-effect transistors," *Appl. Phys. Express*, vol. 7, no. 024201, 2014.

[42] E. A. Casu, W. A. Vitale, N. Oliva, T. Rosca, A. Biswas, C. Alper, A. Krammer, G. V.

Luong, Q. T. Zhao, S. Mantl, A. Schuler, A. Seabaugh, and A. M. Ionescu, "Hybrid phase-change - Tunnel FET (PC-TFET) switch with subthreshold swing < 10mV/decade and sub-0.1 body factor: Digital and analog benchmarking," *Proc. IEDM Tech. Dig.*, pp. 508, 2016.

[43] J. Zhou, G. Han, Y. Peng, Y. Liu, J. Zhang, Q.-Q. Sun, D. W. Zhang, and Y. Hao, "Ferroelectric negative capacitance GeSn PFETs with sub-20 mV/decade subthreshold swing," *IEEE Electron Dev. Lett.*, vol. 38, pp. 1157, 2017.

[44] C. Liu, H.-H. Chen, C.-Ch. Hsu, C.-C. Fan, H.-H. Hsu, and C.-H. Cheng, "Negative capacitance CMOS field-effect transistors with non-hysteretic steep sub-60mV/dec swing and defect-passivated multidomain switching," *Proc. VLSI Symp. Tech. Dig.*, pp. 224, 2019.

## Acknowledgments

This work was mainly supported by the National Science Foundation (NSF) under Award ECCS-1944095, and partially supported by the New York State Energy Research and Development Authority (NYSERDA) under Award 138126 and the New York State Center of Excellence in Materials Informatics (CMI) under Award C160186. The authors acknowledge support from the Vice President for Research and Economic Development (VPRED) at the University at Buffalo. A.C. acknowledges support from the Presidential Fellowship at the University at Buffalo.

## Author contributions

F.Y. and H.L. conceived and supervised the project. M.L. and H.N.J. performed the device fabrication and measurement. M.L., S.S., A.C., and H.L. participated in the sample preparation.

S.W., Y.F., and C.C. synthesized the monolayer $MoS_2$ and performed the material characterization.

X.L., C.Y., Y.L., and Y.H.L. participated in the data analysis.

## Ethics declarations

The authors declare no competing interests.

# Figures

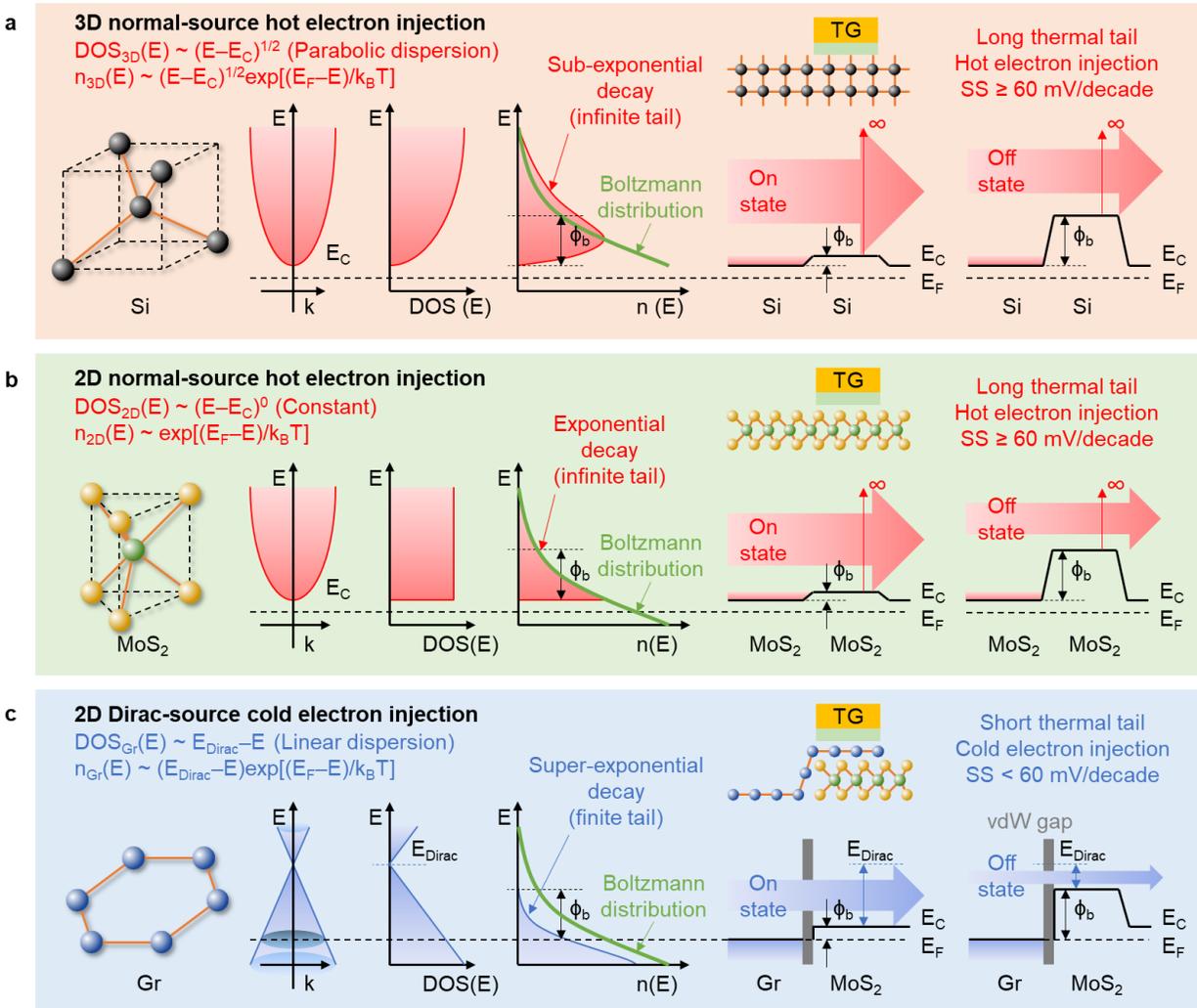

**Fig. 1 Qualitative comparison of normal-source hot electron injection and Dirac-source cold electron injection.** The atomic crystal structure and energy band diagrams including $E$-$k$, $E$-DOS, and $E$-$n$ diagrams are illustrated for (a) 3D semiconductors (e.g., Si), (b) 2D semiconductors (e.g., monolayer $MoS_2$), and (c) 2D monolayer Gr, where $k$ is the wave vector. The corresponding energy band diagrams at the on- and off-state for transistor operation are also included. In the conventional Si or $MoS_2$ FETs, the 3D or 2D normal-source hot electron injection occurs along the in-plane direction and possesses an infinitely long thermal tail over a potential barrier. In the $MoS_2$ DSFET, the 2D Dirac-source cold electron injection has a short thermal tail terminated at $E_{Dirac}$, and can be cut off more efficiently by a gate-controlled $\phi_b$ along the out-of-plane direction.

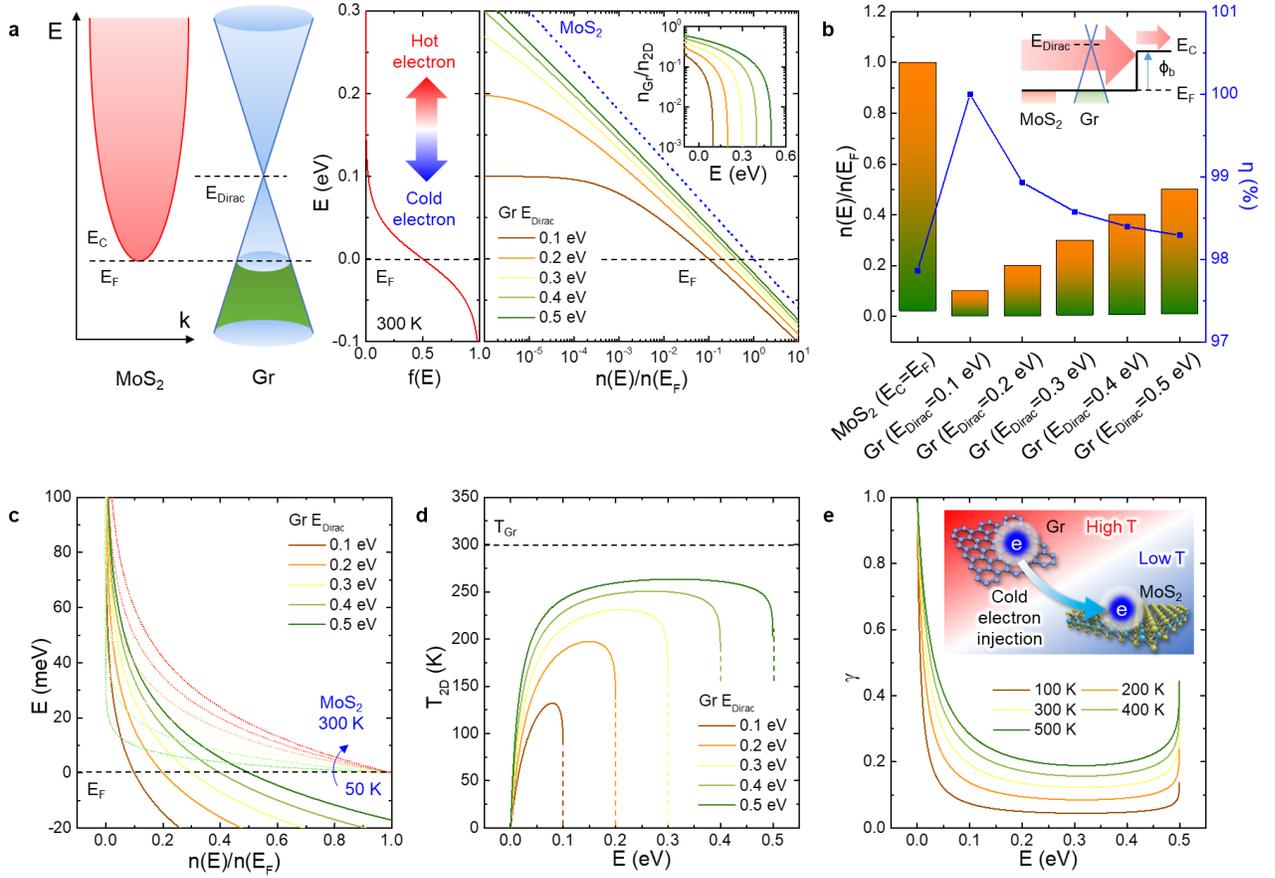

**Fig. 2 Quantitative comparison of 2D normal-source hot electron injection in monolayer MoS$_2$ and 2D Dirac-source cold electron injection in monolayer Gr.** (a) Left: Energy band structures of MoS$_2$ and Gr in $E$-$k$ diagram. Right: Calculated $E$ as a function of $f(E)$ and $n(E)/n(E_F)$. Here $E_F = 0$ eV for both MoS$_2$ and Gr. Inset: $n_{Gr}(E)/n_{2D}(E)$ as a function of $E$ for different doping levels in Gr. (b) The reduction of $n(E)/n(E_F)$ and the corresponding $\eta$ for both MoS$_2$ and Gr when $\phi_b$ increases from 0 to 0.1 eV. Here $E_F = E_C = 0$ eV for MoS$_2$ and $E_F = 0$ eV for Gr. Inset: Energy band structure illustrate the electron flow over $\phi_b$ for MoS$_2$ or Gr. (c) Comparison of the electron excitation of Gr at room temperature (solid lines) and that of MoS$_2$ at low temperature (dot lines, from 50 to 300 K with a step of 50 K). (d) $T_{2D}$ as a function of $E$ when the cold electron injection occurs at $T_{Gr} = 300$ K. Dash lines are guides for the eye. (e) $\gamma$ as a function of $E$ when the cold electron injection occurs at $T_{Gr}$ ranging from 100 to 500 K. Inset: Illustration of an electronic cooling effect when the cold electrons are injected from Gr to MoS$_2$.

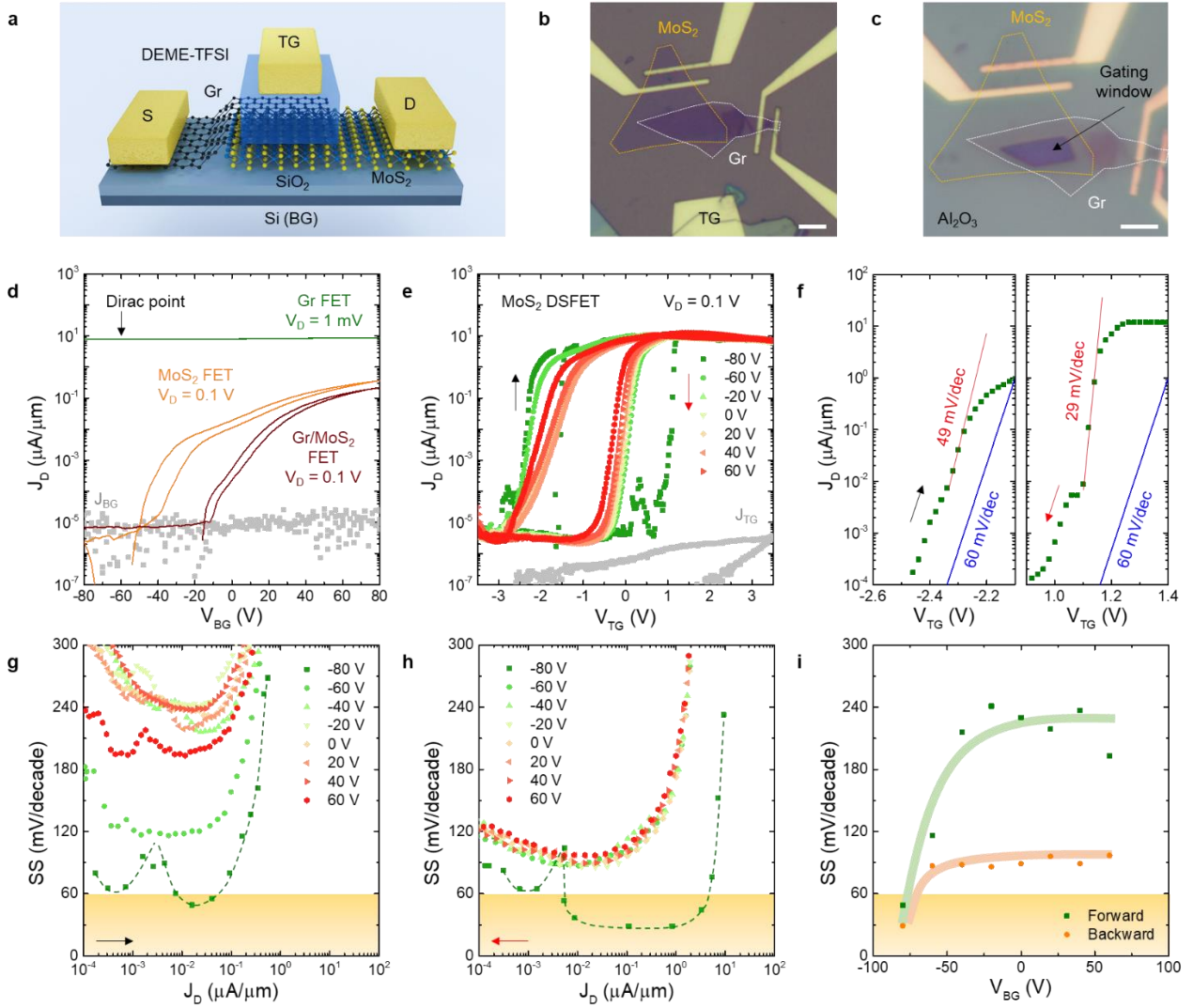

**Fig. 3 Sub-60-mV/decade switching of 2D MoS$_2$ DSFET at room temperature.** (a) Schematic illustration of the device structure. Here S, D, TG, and BG denote the source, drain, top gate, and back gate, respectively. (b, c) Optical microscopy images of the device before and after Al$_2$O$_3$ deposition. An open gating window is opened within the Gr/MoS$_2$ overlapping area for a localized top gating through an ionic liquid. Scale bar: 5 μm. (d) $J_D$-$V_{BG}$ transfer characteristics of Gr FET, MoS$_2$ FET, and Gr/MoS$_2$ FET. (e) $J_D$-$V_{TG}$ transfer characteristics of MoS$_2$ DSFET under various $V_{BG}$. $J_{BG}$ and $J_{TG}$ are the leakage current density measured from the back gate and the top gate, respectively. Back and red arrows indicate the forward and backward sweeps, respectively. (f) Sub-60-mV/decade switching of MoS$_2$ DSFET in the forward and backward sweeps. Blue line is the 60-mV/decade thermionic limit. (g, h) SS as a function of $J_D$ in the forward and backward sweeps under various $V_{BG}$. Dash lines as guides for the eye indicate the double minima of SS at $V_{BG} = -80$ V. (i) The minimum SS as a function of $V_{BG}$ in the forward and backward sweeps with lines as guides for the eye.

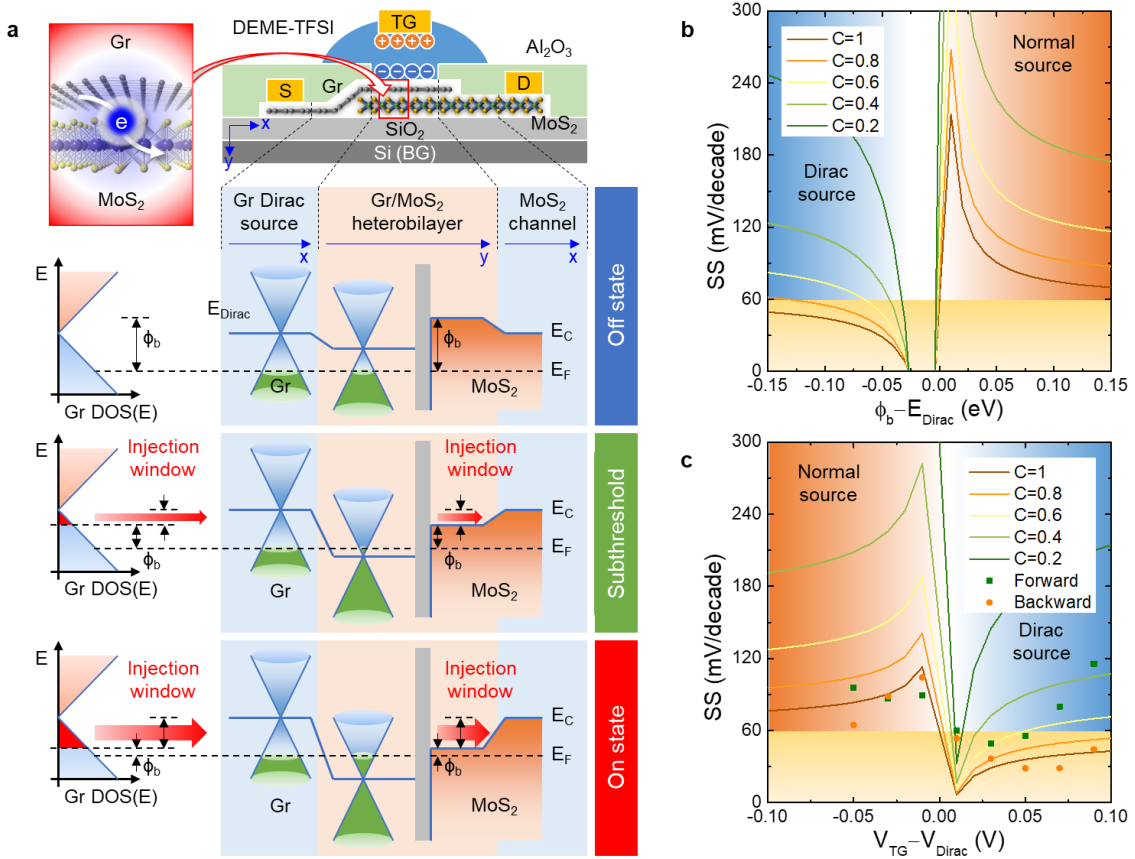

**Fig. 4 Operation principle of 2D MoS$_2$ DSFET.** (a) Cross-section view of the device structure and the energy band diagrams along the carrier transport path for different conditions including the off state, subthreshold (dominated by the Dirac-source cold electron injection), and on state. (b) Calculated SS as a function of $\phi_b$–$E_{Dirac}$ at different $C$ levels. (c) Calculated SS as a function of $V_{TG}$–$V_{Dirac}$ compared with the experimental results.

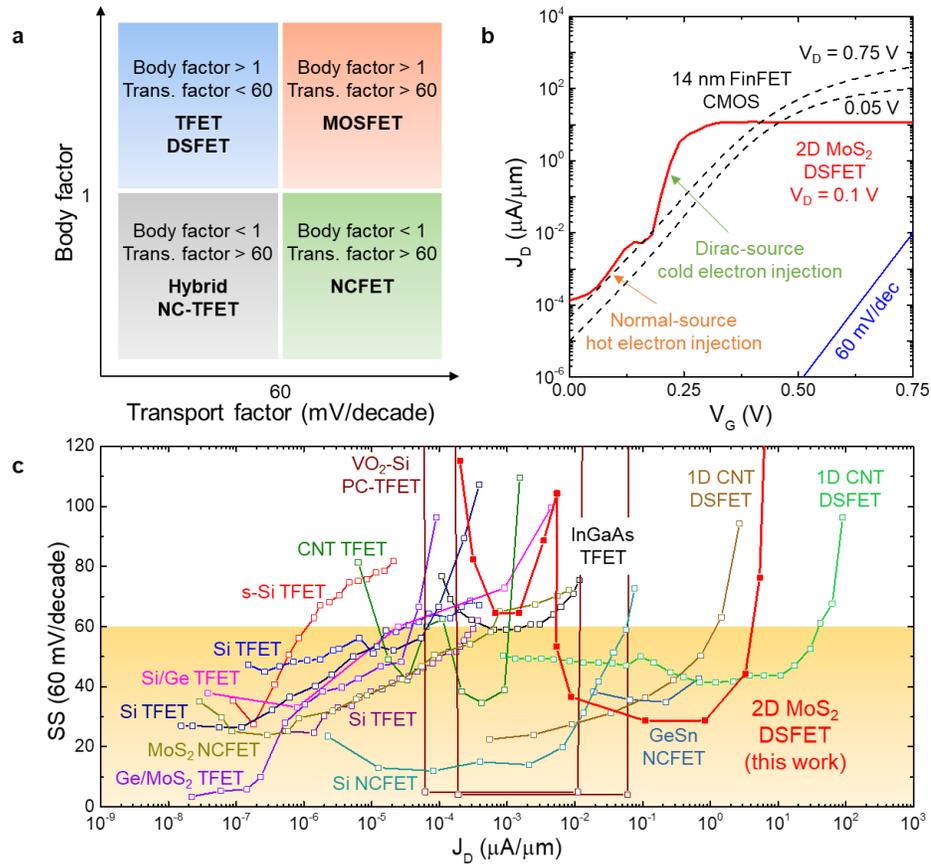

**Fig. 5 Benchmarking of 2D MoS₂ DSFET with other state-of-the-art steep-slope transistor technologies.** (a) Body factor versus transport factor for comparing the beyond-CMOS technologies including TFET, DSFET, NCFET, and hybrid NC-TFET. (a) Transfer characteristics of 2D MoS$_2$ DSFET in comparison with 14 nm FinFET CMOS technology. (b) SS as a function of $J_D$ in a comparison with TFETs, NCFETs, and 1D DSFETs based on a variety of channel materials.

# Supporting Information

# Two-dimensional Cold Electron Transport for Steep-slope Transistors

Maomao Liu,[1] Hemendra Nath Jaiswal,[1] Simran Shahi,[1] Sichen Wei,[2] Yu Fu,[2] Chaoran Chang,[2] Anindita Chakravarty,[1] Xiaochi Liu,[3] Cheng Yang,[4] Yanpeng Liu,[5] Young Hee Lee,[6] Fei Yao,[2,*] and Huamin Li[1,*]

[1] *Department of Electrical Engineering, University at Buffalo, The State University of New York, Buffalo, New York 14260, US*

[2] *Department of Materials Design and Innovation, University at Buffalo, The State University of New York, Buffalo, New York 14260, US*

[3] *School of Physics and Electronics, Central South University, Changsha 410083, China*

[4] *School of Physics and Electronics, Shandong Normal University, Jinan 250014, China*

[5] *Institute of Nanoscience, Nanjing University of Aeronautics and Astronautics, Nanjing 210016, China*

[6] *Center for Integrated Nanostructure Physics, Institute for Basic Science, Suwon 16419, Korea*

\* Author to whom correspondence should be addressed. Electronic address: feiyao@buffalo.edu and huaminli@buffalo.edu

## 1. Material characterization

The Raman spectroscopy was performed by Renishaw inVia Raman microscope. The atomic force microscopy (AFM) was performed by Bruker Dimension Icon with ScanAsyst. Both the Raman and AFM characterization confirmed the monolayer structure of $MoS_2$ and Gr, as shown in **Fig. S1**. Specifically, the Raman spectrum of the monolayer $MoS_2$ shows an $E^1_{2g}$ peak at 385 cm$^{-1}$ and an $A_{1g}$ peak at 402 cm$^{-1}$. The AFM characterization of the monolayer $MoS_2$ suggests a thickness of ~0.7 nm. For the monolayer Gr, a *2D* peak at 2,687 cm$^{-1}$ and a *G* peak at 1,587 cm$^{-1}$ are obtained in the Raman spectrum, and the *2D/G* peak intensity ratio is about 2, serving as the signature of the monolayer Gr. The AFM characterization of the monolayer Gr suggests a thickness of ~0.6 nm. Compared to the ideal thickness value (0.34 nm for the monolayer Gr and 0.65 nm for the monolayer $MoS_2$), the difference in this work is mainly due to the polymer residues which are induced on the sample surface during the transfer and lithography process. By combining both the Raman and AFM characterizations, we can conclude the monolayer structure of $MoS_2$ and Gr.

## 2. Device fabrication and measurement

The monolayer $MoS_2$ was synthesized using a customized two-zone chemical vapor deposition (CVD) system [S1]. Specifically, ammonium heptamolybdate was used as a water-soluble Mo precursor, and NaOH was used as a water-soluble promoter. Their mixed solution was spin-coated on a $SiO_2$/Si growth substrate. The reaction between Mo and Na produced $Na_2MoO_4$ compounds and then become $MoS_2$ after S vapor injection. The annealing time was optimized to control the size of the isolated monolayer triangular domains. After synthesis, the monolayer $MoS_2$ flakes were wet-transferred onto an n-type Si substrate (0.001-0.005 Ωcm) with a 285-nm-thick

SiO$_2$ layer on the surface. Then, the monolayer graphene (Gr) was mechanically exfoliated from graphite crystal and transferred on top of the monolayer MoS$_2$ to form a partially overlapped heterobilayer structure. Next, multiple Ti/Au electrodes (10 nm/90 nm) were patterned and deposited using electron-beam lithography and evaporation, serving as the source, drain, and top gate for different transistor configurations including the Gr FET, MoS$_2$ FET, Gr/MoS$_2$ FET, and MoS$_2$ DSFET, as shown in **Fig. S2**. Before applying the ionic liquid, a 30-nm-thick Al$_2$O$_3$ thin film was deposited by atomic layer deposition (ALD), followed by electron-beam lithography and wet etching to create two well-confined gating windows: one on the heterobilayer and another one on the top gate electrode. After drop-casting DEME-TFSI (727679 Sigma-Aldrich), the gating windows were connected through the ionic liquid, and a localized top gate voltage can be applied to the exposed Gr/MoS$_2$ heterobilayer through the electrical double-layer (EDL) effect.

The electrical measurements were performed in a vacuum-chamber probe station (MSTECH M5VC) with a semiconductor parameter analyzer (Keysight B1500A), and the temperature ($T$) varied from 295 to 358 K. The drain current ($I_D$) was measured as a function of the back-gate voltage ($V_{BG}$), tog-gate voltage ($V_{TG}$), and drain voltage ($V_D$) for the output and transfer characteristics. The top-gate sweep rate was set as 1 mV/s to ensure the EDL establishment as $V_{TG}$ varies [S2-S4]. Due to the triangle shape of the monolayer MoS$_2$ domain and thus the irregular channel structure, the drain current density ($J_D$) was defined as $I_D$ divided by the length of the drain contact electrode on the MoS$_2$ layer.

It is also worth mentioning that the leakage currents through the back gate ($J_{BG}$) and the top gate ($J_{TG}$) were well-maintained at around or even less than $10^{-5}$ μA/μm, as shown in **Fig. 3(d)** and **(e)**. Therefore, the measured $J_D$ values which were higher than $J_{BG}$ and $J_{TG}$ were reliable for all the devices in further characterizations.

## 3. $J_D$-$V_D$ output characteristics

A comparison of the $J_D$-$V_D$ output characteristics among the Gr FET, MoS$_2$ FET, Gr/MoS$_2$ FET, and MoS$_2$ DSFET is carried out at different $V_{BG}$ or $V_{TG}$, as shown in **Fig. S3**. Specifically, $J_D$ in the Gr FET shows a narrow variation with $V_{BG}$, due to a weak gate modulation effect. The MoS$_2$ FET and Gr/MoS$_2$ FETs show a monotonic increase of $J_D$ with $V_{BG}$, clearly suggesting an electron transport in n-type FETs. The MoS$_2$ DSFET also shows a monotonic increase of $J_D$ as a function of $V_{TG}$ at $V_{BG} = -80$ V, yet a dramatic jump of $J_D$ occurs when $V_{TG}$ increases from 0 to 1 V. Then $J_D$ remains as a constant even $V_{TG}$ increases further to 2 V. These results are consistent with the features of the DSFET, i.e., the steep-slope transition between off and on states as well as the strong current saturation at the on state.

It is also worth mentioning that all the devices show a clear linear $JV$ relationship. This result indicates an Ohmic contact condition achieved for all the devices in this work.

## 4. $J_D$-$V_{BG}$ transfer characteristics

A comparison of the $J_D$-$V_{BG}$ transfer characteristics among the Gr FET, MoS$_2$ FET, and Gr/MoS2 FET is performed, as shown in **Fig. 3(d)**. The figure is then replotted in a linear scale, as shown in **Fig. S4**, in order to extract the corresponding threshold voltage ($V_{th}$) and the field-effect mobility ($\mu_{FE}$). Here $\mu_{FE}$ is defined as $L(1/C_{ox})(1/V_D)g_m$, where $L$ is the MoS$_2$ channel length defined as the shortest distance between the Gr/MoS$_2$ heterobilayer area and the drain contact electrode on the MoS$_2$ layer; $C_{ox}$ is the oxide capacitance ($1.2\times10^{-8}$ F/cm$^2$) for a 285-nm-thick SiO$_2$ layer; $g_m$ is the transconductance defined as $\partial J_D/\partial V_G$ at certain $V_D$. At room temperature, the Gr FET shows the Dirac point at about $-60$ V of $V_{BG}$ and the maximum $\mu_{FE}$ of 4,957 cm$^2$/Vs for electrons; the MoS$_2$ FET shows $V_{th}$ at about 20 V of $V_{BG}$ and the maximum $\mu_{FE}$ of 24 cm$^2$/Vs for electrons. The Gr/MoS$_2$ FET possesses $V_{th}$ at about 20 V as well. The maximum $\mu_{FE}$ is 66 cm$^2$/Vs for electrons which is an intermediate value compared to the Gr FET and the MoS$_2$ FET. This result is in a good agreement with the previous report which has a similar device structure [S5].

## 5. Transconductance characterizations

A comparison of $g_m$ among Gr FET, MoS$_2$ FET, Gr/MoS$_2$ FET, and MoS$_2$ DSFET is performed to investigate the intrinsic gain and switching speed, as shown in **Fig. S5**. The conventional Gr FET, MoS$_2$ FET, and Gr/MoS$_2$ FET show the maximum $g_m$ at around 10$^{-2}$ μS/μm, whereas the MoS$_2$ DSFET has the maximum $g_m$ at around 10 μS/μm, suggesting a significant improvement. The transconductance efficiency, defined as $g_m/J_D$, is also calculated for all the devices. The MoS$_2$ DSFET has the maximum $g_m/J_D$ over 400 V$^{-1}$, which is about one order of the magnitude higher than the limit (38.5 V$^{-1}$) in the conventional transistors. Such superior performance is also comparable with the recent report on 1D CNT DSFETs [S6].

## 6. Carrier mobility in MoS$_2$ DSFET

The MoS$_2$ DSFET is operated through an EDL gating using the ionic liquid. The calculation of $\mu_{FE}$ in this case is challenging, as the setup for measuring EDL capacitance ($C_{EDL}$) is complicated [S7]. Even using the same ionic liquid, the actual $C_{EDL}$ might vary significantly due to the difference in experimental setups (e.g., sweep rate, temperature, and humidity) and sample conditions (e.g., device structure, surface roughness and contamination). Therefore, we exploit the extreme cases of $C_{EDL}$ to anticipate the possible range of $\mu_{FE}$ in the MoS$_2$ DSFET. In general, $C_{EDL}$ ranges from 1 to 100 μF/cm$^2$ [S7], which is orders of magnitude higher than the conventional oxide capacitance, e.g., $1.2 \times 10^{-2}$ μF/cm$^2$ for a 285-nm-thick SiO$_2$ layer or $3.8 \times 10^{-2}$ μF/cm$^2$ for a 90-nm-thick SiO$_2$ layer. Here we choose three $C_{EDL}$ values (1, 10, and 100 μF/cm$^2$) and calculate the corresponding $μ_{FE}$ defined as $L(1/C_{EDL})(1/V_D)g_m$ for electrons in the MoS$_2$ DSFET, as shown in **Fig. S6**. The estimated maximum $μ_{FE}$ ranges from 6 to 573 cm$^2$/Vs in the forward sweep and from 86 to 8,614 cm$^2$/Vs in the backward sweep. Considering the predicted room-temperature, phonon-limited (also known as "intrinsic") electron mobility of the monolayer MoS$_2$ is up to 480 cm$^2$/Vs [S8-S11], the corresponding $C_{EDL}$ can be calculated to be about 18 μF/cm$^2$. This result is in a good agreement with the previous report on $C_{EDL}$ using DEME-TFSI applied on MoS$_2$ [S12].

## 7. Threshold voltages and hysteresis in MoS$_2$ DSFET

For the MoS$_2$ DSFET, the value of $V_{th}$ as a function of $V_{BG}$ ranging from –80 to 60 V was obtained at $V_D = 0.1$ V by replotting **Fig. 3(e)** in linear scale, as shown in **Fig. S7**. As $V_{BG}$ increases, $V_{th}$ in the forward sweep shifts toward the negative $V_{BG}$ direction, and $V_{th}$ in the backward sweep shifts toward the positive $V_{BG}$ direction. As a result, the hysteresis window ($ΔV_{th}$), defined as the difference of $V_{th}$ between the forward and backward sweeps, is obtained as a function of $V_{BG}$. The extracted $ΔV_{th}$ shows a monotonic decrease from 3.5 to 0.5 V as $V_{BG}$ increases from –80 to 60 V. In principle, the hysteresis is attributed to the intrinsic defects (such as S vacancies) in MoS$_2$ which act as charge traps with a strong dependence on the electrostatic gating, as well as H$_2$O and O$_2$ absorbates on the unpassivated device surface [S13]. Because our device is passivated with the thin Al$_2$O$_3$ layer and the ionic liquid, and measured in a vacuum environment, the possible foreign absorbates such as H$_2$O and O$_2$ can be excluded. The strong dependence of $ΔV_{th}$ on $V_{BG}$ further confirms the dominating role of the intrinsic defects for the hysteresis in MoS$_2$. This result is consistent with the previous report on the hysteresis observed in the MoS$_2$ FET [S13].

## 8. Reproducibility

To confirm the reproducibility, we also fabricate additional MoS$_2$ DSFETs with a similar device structure as shown in **Fig. 3**. Optical microscopy images of one example device are shown in **Fig. S8(a)** and **(b)**. A summary of these device performances is shown in **Fig. S8(c)** and **(d)**. From the room-temperature $J_D$-$V_{TG}$ characteristics, all the devices clearly show a strong on-current saturation across a wide $V_{TG}$ range (~5 V). The extracted SS is plotted as a function of $J_D$. Some of the devices didn't possess a sub-60-mV/decade SS, probably due to the contamination at the Gr/MoS$_2$ interface which was induced during the transfer process [S14]. Nevertheless, a clear "double-minima" feature is obtained for all the devices. Specifically, the second SS minima locates at a higher $J_D$ and corresponds to the Dirac-source carrier injection. Its value is always lower than the first SS minima which locates at a lower $J_D$ and corresponds to the normal-source carrier injection. In summary, both the strong on-current saturation and the unique double-minima SS are obtained in multiple MoS$_2$ DSFETs, in addition to the sub-60-mV/decade SS. These results indicate the reproducibility and reliability of the MoS$_2$ DSFETs in this work.

## 9. Temperature-dependent measurement

We fabricate another MoS$_2$ DSFET to perform a temperature ($T$)-dependent measurement, as shown in **Fig. S9**. Because DEME-TFSI is a room-temperature ionic liquid, ion mobility would decrease significantly as $T$ decreases, even $T$ is still higher than the glass transition temperature. Therefore, we choose a high-$T$ range near and above the room temperature for this measurement. The highest $T$ in this work is set as 358 K, in order to prevent possible failure or damage of the ionic liquid at the high $T$. We also closely monitor the $J_{TG}$ variations at the high $T$, and our results confirm that $J_{TG}$ was still well-controlled at around $10^{-5}$ µA/µm at 358 K.

For this specific DSFET, the extracted SS doesn't show a clear sub-60-mV/decade value. Nevertheless, the extracted SS still shows a clear increase as a function of $T$, suggesting the dominance of thermionic emission in carrier transport. This result is consistent with our theory and previous reports regarding the Dirac source carrier injection for CNT channels [S15]. We also extract two current values and plot them as a function of $T$. One is the on-state $J_D$ ($J_{D,on}$) obtained at $V_{TG}$ = 3.5 V. Another one is $J_D$ corresponding to the minimum SS ($J_{D,SSmin}$). Both $J_{D,on}$ and $J_{D,SSmin}$ show a clear increase with $T$, being consistent with the thermionic carrier transport behavior.

## 10. Theoretical calculations

Both normal-source and Dirac-source carrier injection are dominated by thermionic emission, their current can be described by the Landauer-Büttiker formula at the ballistic transport limit [S14-S16] as

$$I_{thermal} = \frac{2q}{h} \int P(E) D(E) \left[ f(E - E_{FS}) - f(E - E_{FD}) \right] dE, \quad (1)$$

where $P(E)$ is the transmission probability, $D(E)$ is the DOS related to Gr and $MoS_2$, $E_{FS}$ and $E_{FD}$ are $E_F$ at source and drain, respectively. Because the DOS of Gr is a linear function of $E$ near $E_{Dirac}$ and the DOS of $MoS_2$ is constant, the DOS of the Gr/$MoS_2$ heterobilayer is described as $D(E) = D_0|E - E_{Dirac}|$ where $D_0$ is constant. When $E_{FS} < \phi_b$, the current is determined by the thermal tail of the Fermi-Dirac distribution function above $E_{FS}$, and thus an approximation can be made as $f(E-E_{FS}) - f(E-E_{FD}) \approx \exp(-E/k_B T)$. To anticipate the upper-limit performance, we assume a unity $P(E)$, and the SS can be obtained as

$$SS = \frac{\partial V_{TG}}{\partial \log_{10}(I_{thermal})} = \frac{k_B T \ln 10}{q} \frac{1}{C} \left(1 + \frac{k_B T}{\phi_b - E_{Dirac}}\right), \quad (2)$$

where $C = \partial \phi_b / \partial(qV_{TG})$ is the reciprocal body factor ($0 < C < 1$). Thus, the SS can be plotted as a function of $\phi_b - E_{Dirac}$ (see **Fig. 4(b)**).

Next, we assume the EDL capacitance in the ionic liquid is 1 µF/cm$^2$ [S12], and calculate the Fermi level shift ($E_F - E_{Dirac}$) of Gr under the top gating as $E_F - E_{Dirac} = \hbar v_F (\pi \alpha |V_{TG} - V_{Dirac}|)^{1/2}$, where $\hbar$ is the reduced Planck's constant, $v_F$ is the Fermi velocity in Gr (~10$^6$ m/s), and $\alpha$ is the electrostatic gate coupling factor (~6.3×10$^{12}$ cm$^{-2}$ V$^{-1}$ for 1 µF/cm$^2$) [S17]. Thus, the SS as a function of $V_{TG} - V_{Dirac}$ can be extracted (see **Fig. 4(c)**). As $V_{TG}$ increases, the normal-source and Dirac-source carrier injections are predicted in succession, which are qualitatively consistent with the experimental data obtained in both forward and backward sweeps.

# References

[S1] H. Kim, G. H. Han, S. J. Yun, J. Zhao, D. H. Keum, H. Y. Jeong, T. H. Ly, Y. Jin, J. H. Park, B. H. Moon, S. W. Kim, and Y. H. Lee, "Role of alkali metal promoter in enhancing lateral growth of monolayer transition metal dichalcogenides," *Nanotechnology*, vol. 28, 36LT01, 2017.

[S2] H. Li, K. Xu, B. Bourdon, H. Lu, Y.-C. Lin, J. A. Robinson, A. C. Seabaugh, and S. K. Fullerton-Shirey, "Electric double layer dynamics in poly(ethylene oxide) LiClO$_4$ on graphene transistors," *J. Phys. Chem. C*, vol. 121, pp. 16996, 2017.

[S3] X. Liu, D. Qu, H. Li, I. Moon, F. Ahmed, C. Kim, M. Lee, Y. Choi, J. H. Cho, J. C. Hone, and W. J. Yoo, "Modulation of quantum tunneling via a vertical two-dimensional black phosphorus and molybdenum disulfide p-n junction," *ACS Nano*, vol. 11, pp. 9143, 2017.

[S4] C. Alessandri, S. Fathipour, H. Li, I. Kwak, A. Kummel, M. Remskar, and A. Seabaugh, "Reconfigurable electronic double layer doping in an MoS$_2$ nanoribbon transistor," *IEEE Trans. Electron Devices*, vol. 64, pp. 5217, 2017.

[S5] H. Tian, Z. Tan, C. Wu, X. Wang, M. A. Mohammad, D. Xie, Y. Yang, J. Wang, L.-J. Li, J. Xu, and T.-L. Ren, "Novel field-effect Schottky barrier transistors based on graphene-MoS$_2$ heterojunctions," *Sci. Rep.*, vol. 4, no. 5951, 2015.

[S6] L. Xu, C. Qiu, L.-M. Peng, and Z. Zhang, "Transconductance amplification in Dirac-source field-effect transistors enabled by graphene/nanotube heterojunctions," *Adv. Electron. Mater.*, vol. 6, no. 1901289, 2020.

[S7] H. Ji, X. Zhao, Z. Qiao, J. Jung, Y. Zhu, Y. Lu, L. L. Zhang, A. H. MacDonald, and R. S. Ruoff, "Capacitance of carbon-based electrical double-layer capacitors," *Nat. Commun.*, vol. 5, no. 3317, 2014.

[S8] K. Kaasbjerg, K. S. Thygesen, and K. W. Jacobsen, "Phonon-limited mobility in n-type single-layer MoS$_2$ from first principles," *Phys. Rev. B*, vol. 85, no. 115317, 2012.

[S9] X. Li, J. T. Mullen, Z. Jin, K. M. Borysenko, M. B. Nardelli, and K. W. Kim, "Intrinsic electrical transport properties of monolayer silicene and MoS$_2$ from first principles," *Phys. Rev. B*, vol. 87, no. 115418, 2013.

[S10] Z. Jin, X. Li, J. T. Mullen, and K. W. Kim, "Intrinsic transport properties of electrons and holes in monolayer transition-metal dichalcogenides," *Phys. Rev. B*, vol. 90, no. 045422, 2014.

[S11] T. Gunst, T. Markussen, K. Stokbro, and M. Brandbyge, "First-principles method for electron-phonon coupling and electron mobility: Applications to two-dimensional materials," *Phys. Rev. B*, vol. 93, no. 035414, 2016.
35

[S12] Y. Zhang, J. Ye, Y. Matsuhashi, and Y. Iwasa, "Ambipolar MoS$_2$ thin flake transistors," *Nano Lett.*, vol. 12, pp. 1136, 2012.

[S13] A. Di Bartolomeo, L. Genovese, F. Giubileo, L. Iemmo, G. Luongo, T. Foller, and M. Schleberger, "Hysteresis in the transfer characteristics of MoS$_2$ transistors," *2D Mater.*, vol. 5, no. 1, 2017.

[S14] F. Liu, C. Qiu, Z. Zhang, L.-M. Peng, J. Wang, and H. Guo, "Dirac electrons at the source: Breaking the 60-mV/decade switching limit," *IEEE Trans. Electron Devices*, vol. 65, pp. 2736, 2018.

[S15] C. Qiu, F. Liu, L. Xu, B. Deng, M. Xiao, J. Si, L. Lin, Z. Zhang, J. Wang, H. Guo, H. Peng, and L.-M. Peng, "Dirac-source field-effect transistors as energy-efficient, high-performance electronic switches," *Science*, vol. 361, pp. 387, 2018.

[S16] L. Britnell, R. V. Gorbachev, R. Jalil, B. D. Belle, F. Schedin, A. Mishchenko, T. Georgiou, M. I. Katsnelson, L. Eaves, S. V. Morozov, N. M. R. Peres, J. Leist, A. K. Geim, K. S. Novoselov, and L. A. Ponomarenko, "Field-effect tunneling transistor based on vertical graphene heterostructures," *Science*, vol. 335, pp. 947, 2012.

[S17] F. Wang, Y. Zhang, C. Tian, C. Girit, A. Zettl, M. Crommie, and Y. R. Shen, "Gate-variable optical transitions in graphene," *Science*, vol. 320, pp. 206, 2008.

# Figures

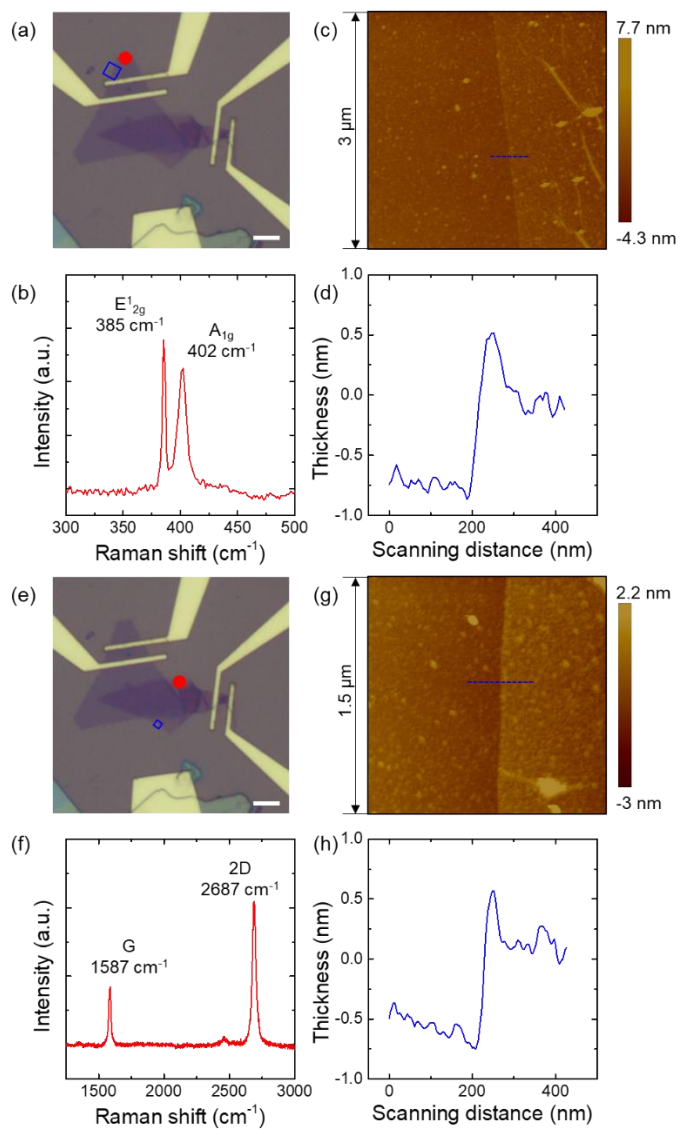

**Fig. S1** (a-d) Characterization of monolayer $MoS_2$ in the 2D $MoS_2$ DSFET, including microscopy image, Raman spectrum, AFM surface scan and height profile. Blue box indicates the AFM scan area. Red dot indicates the Raman laser spot. Scale bar: 5 μm. (e-h) Characterization of monolayer Gr in the 2D $MoS_2$ DSFET, including microscopy image, Raman spectrum, AFM surface scan and height profile.

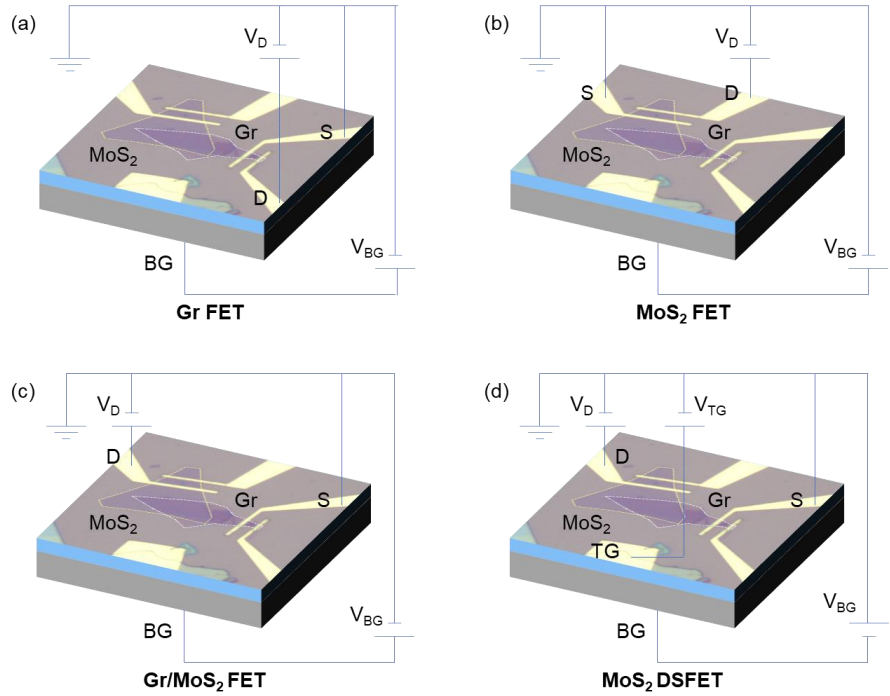

**Fig. S2** Schematic illustration of transistor configurations for (a) Gr FET, (b) MoS$_2$ FET, (c) Gr/MoS$_2$ FET, and (d) MoS$_2$ DSFET. Here S, D, BG, TG denote the source, drain, back gate, and top gate, respectively. $V_D$, $V_{BG}$, and $V_{TG}$ denote the drain voltage, back-gate voltage, and top-gate voltage, respectively.

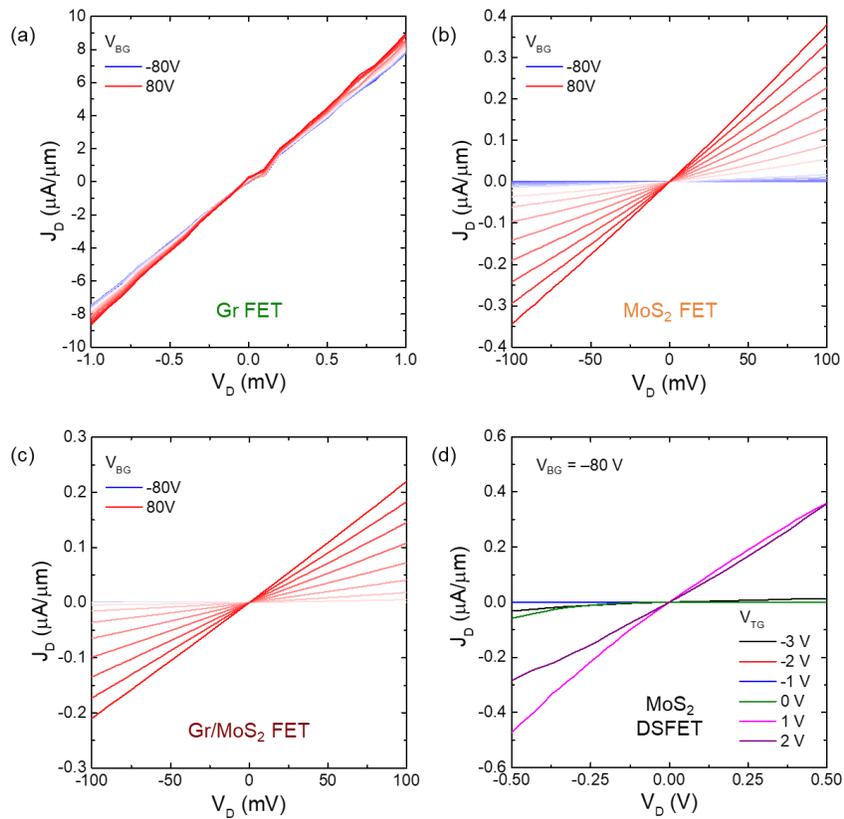

**Fig. S3** Comparison of $J_D$-$V_D$ transfer characteristics of (a) Gr FET, (b) MoS$_2$ FET, and (c) Gr/MoS$_2$ FET under the back gating, as well as (d) MoS$_2$ DSFET under the top gating at $V_{BG}$ = −80 V. The linear $JV$ relationship at small $V_D$ suggests an Ohmic contact for all the devices. An abrupt increase of $J_D$ in the MoS$_2$ DSFET is obtained as $V_{TG}$ switches from 0 to 1 V. Then $J_D$ saturates even $V_{TG}$ continues increasing to 2 V.

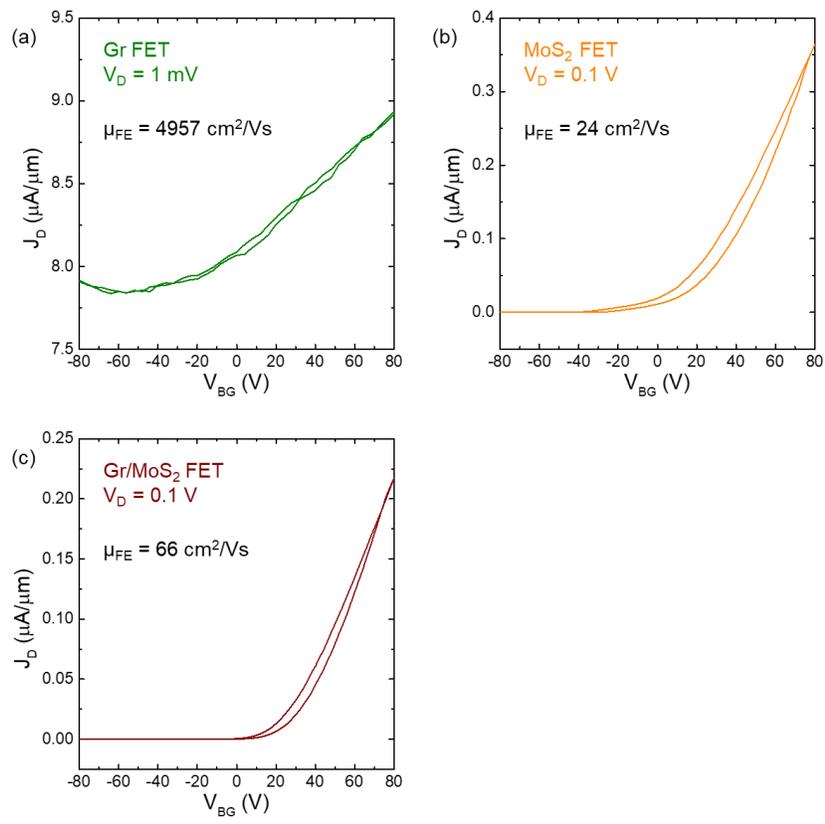

**Fig. S4** $J_D$-$V_{BG}$ transfer characteristics of (a) Gr FET, (b) MoS$_2$ FET, and (c) Gr/MoS$_2$ FET in linear scale, where the Dirac point, $V_{th}$, and maximum $\mu_{FE}$ were extracted.

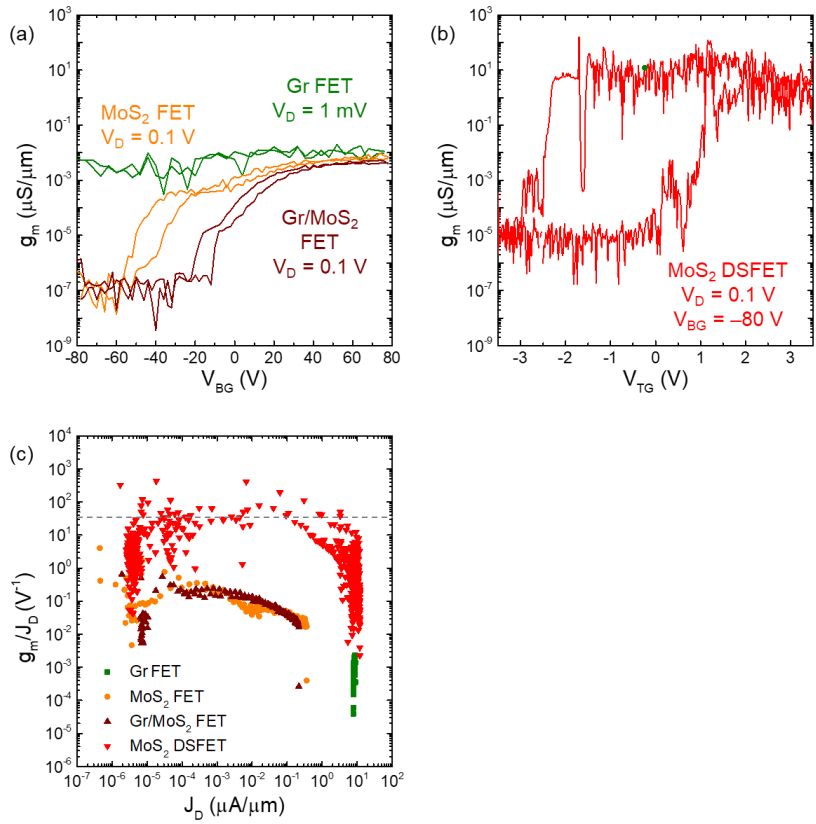

**Fig. S5** (a, b) Comparison of $g_m$ characteristics of Gr FET, MoS$_2$ FET, Gr/MoS$_2$ FET, and MoS$_2$ DSFET. (c) $g_m/J_D$ as a function of $J_D$ illustrates the transconductance efficiency. The gray dash line indicates the efficiency limit (38.5 V$^{-1}$) of the conventional transistors.

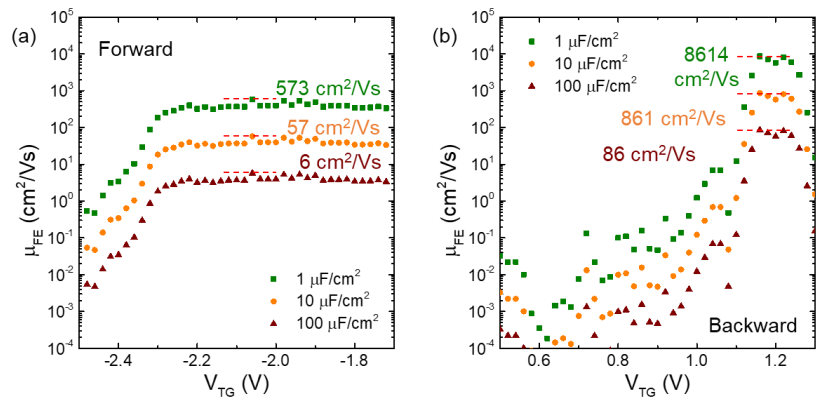

**Fig. S6** Estimation of $\mu_{FE}$ for electrons in MoS$_2$ DSFET in (a) forward sweep and (b) backward sweep at $V_{BG} = -80$ V and $V_D = 0.1$ V. The $C_{EDL}$ value is assumed to be 1, 10, and 100 µF/cm$^2$.

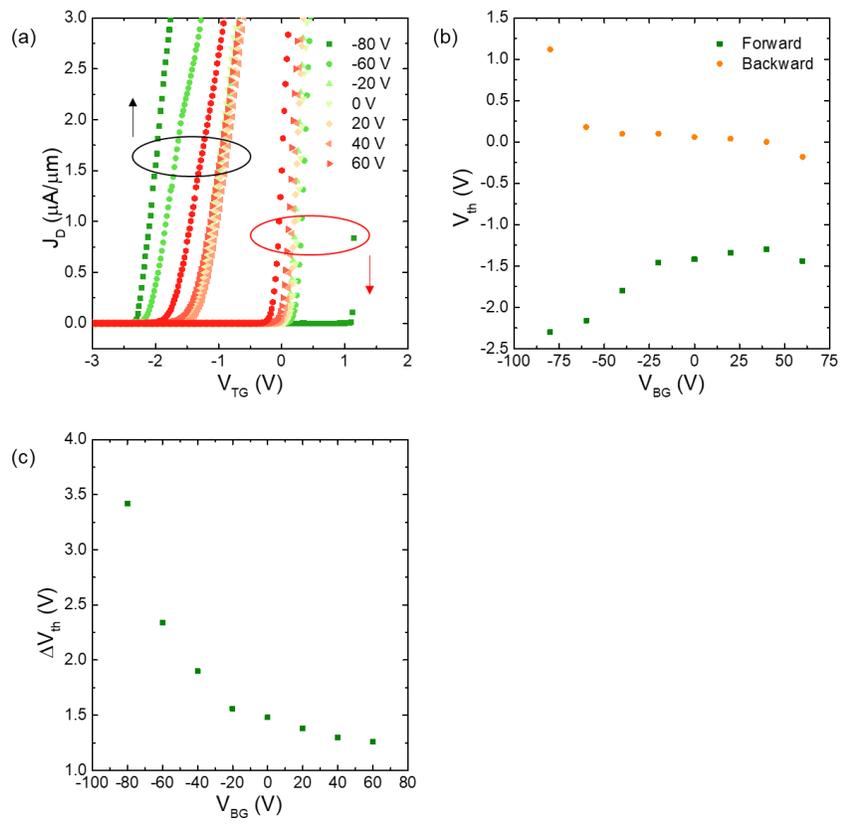

**Fig. S7** (a) $J_D$-$V_{TG}$ transfer characteristics of MoS$_2$ DSFET in linear scale. Both the forward and backward sweeps at $V_D = 0.1$ V are included as a function of $V_{BG}$ ranging from –80 V to 60 V. (b) The extracted $V_{th}$ as a function of $V_{BG}$. (c) The extracted $\Delta V_{th}$ as a function of $V_{BG}$.

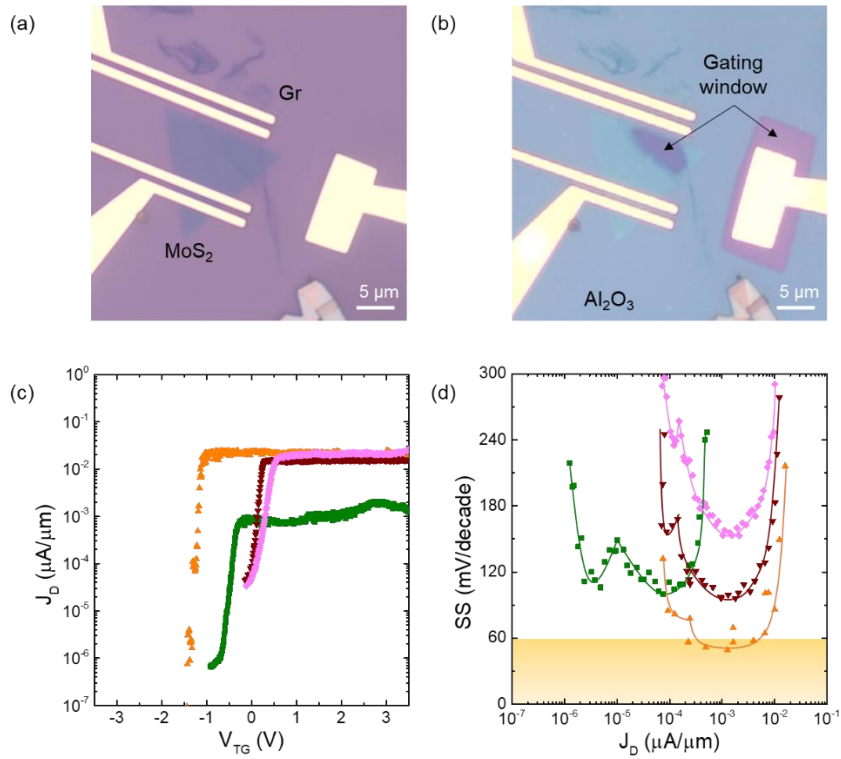

**Fig. S8** A summary of additional MoS$_2$ DSFETs fabricated in a similar device structure. (a, b) Microscopy images of another MoS$_2$ DSFET. Based on a Gr/MoS$_2$ heterobilayer FET structure, a thin layer of Al$_2$O$_3$ was deposited, followed by an opening of the localized gating windows. (c) Room-temperature $J_D$-$V_{TG}$ transfer characteristics at $V_{BG}$ = –80 V and $V_D$ = 0.1 V. The novel and strong on-current saturation is clearly obtained for all the devices. (d) The extracted SS as a function of $J_D$ with the solid lines as guides to the eye. The "double-minima" signature is clearly obtained for all the devices.

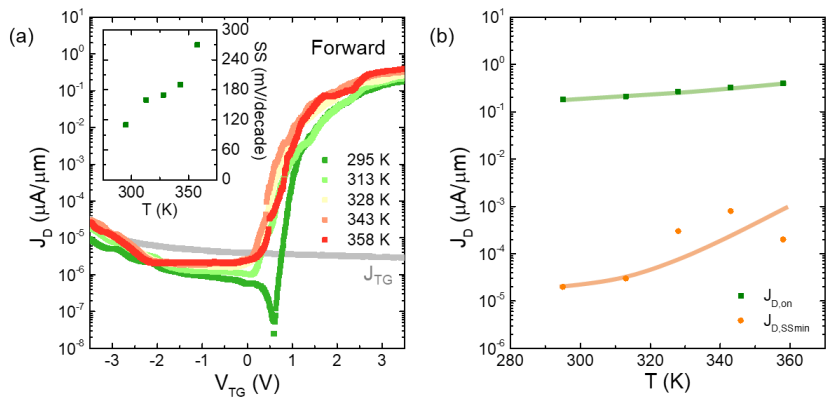

**Fig. S9** (a) The *T*-dependent measurement of $J_D$-$V_{TG}$ transfer characteristics at $V_{BG} = -80$ V and $V_D = 0.1$ V. Inset: The minimum SS value extracted as a function of *T*. (d) The extracted $J_{D,on}$ and $J_{D,SSmin}$ as a function of *T*. The solid lines are guides to the eye.